\documentclass[lettersize,journal]{IEEEtran}
\usepackage{amsmath,amsfonts}
\usepackage{algorithmic}
\usepackage{algorithm}
\usepackage{array}
\usepackage[caption=false,font=normalsize,labelfont=sf,textfont=sf]{subfig}
\usepackage{textcomp}
\usepackage{stfloats}
\usepackage{url}
\usepackage{verbatim}
\usepackage{graphicx}
\usepackage{cite}
\usepackage{algorithmic}
\def\BibTeX{{\rm B\kern-.05em{\sc i\kern-.025em b}\kern-.08em
    T\kern-.1667em\lower.7ex\hbox{E}\kern-.125emX}}
\AtBeginDocument{\definecolor{tmlcncolor}{cmyk}{0.93,0.59,0.15,0.02}\definecolor{NavyBlue}{RGB}{0,86,125}}

\usepackage[colorlinks=true, linkcolor=black, citecolor=black, urlcolor=NavyBlue]{hyperref}

\usepackage{academicons}
\usepackage{xcolor}

\usepackage{gensymb}
\usepackage{enumitem}
\usepackage{amsthm}   
\usepackage{microtype} 
\usepackage{booktabs} 
\usepackage{array}
\newcolumntype{L}[1]{>{\raggedright\arraybackslash}p{#1}}
\newcolumntype{C}[1]{>{\centering\arraybackslash}p{#1}}
\usepackage{tabularx}

\usepackage{amsmath,amssymb}
\usepackage{ragged2e}
\usepackage{xspace} 

\microtypesetup{protrusion=false}

\newcommand{\step}[2]{%
  \par\hangindent=2em
  {\RaggedRight
  \textbf{#1:}\xspace%
  \parbox[t]{0.9\linewidth}{#2}%
  }%
  \par
}

\newcounter{AlgorithmCounter}
\newcommand{\algheader}[1]{%
  \refstepcounter{AlgorithmCounter}%
  \par\hrule height 0.8pt
  \vspace{2pt}%
  \textbf{   Algorithm \theAlgorithmCounter: #1.}%
  \vspace{2pt}\hrule height 0.8pt
  \vspace{6pt}%
}

\usepackage{titlesec}    


\makeatletter
\renewcommand\subsubsection{\@startsection{subsubsection}{3}{0pt}%
  {1em}    
  {0.2em}  
  {\normalfont\normalsize}}
\makeatother

\makeatletter

\let\subsubsubsection\relax

\newcounter{deepsec}[subsubsection]

\renewcommand{\thedeepsec}{\arabic{subsubsection}.\arabic{deepsec}}

\newcommand{\subsubsubsection}[1]{%
  \refstepcounter{deepsec}%
  \par\vspace{0.5\baselineskip}
  \noindent{\thedeepsec\ #1}\par
  \vspace{0.25\baselineskip}
}

\makeatother

\begin{document}

\title{Secure and Parallel Determinant Computation for Large-Scale Matrices in Edge Environments}

\author{Prajwal~Panth,~\IEEEmembership{Student~Member,~IEEE}%
\thanks{Corresponding author: Prajwal Panth (email: prajwal.panth21@gmail.com).}%
\thanks{School of Computer Engineering, KIIT Deemed to be University, Bhubaneswar, India.}%
}

\markboth{}{P.~Panth: Secure and Parallel Determinant Computation for Large-Scale Matrices in Edge Environments}

\maketitle

\begin{abstract}
The advent of edge computing has enabled resource-constrained clients to delegate intensive computational tasks to distributed edge servers, especially within Internet of Things (IoT) environments. Among such tasks, Matrix Determinant Computation (MDC) remains critical for applications in control systems, cryptography, and machine learning. However, the cubic complexity of traditional determinant algorithms makes them unsuitable for real-time processing in constrained edge scenarios.

We propose a Secure Parallel Determinant Computation (SPDC) framework, which provides strong security guaranties, including privacy-preserving MDC, across $N$ distributed edge servers. The framework achieves privacy through Composite Element Distortion (CED)—a lightweight encryption method that combines Element-wise Obfuscation (EWO) and the Panth Rotation Theorem (PRT) to conceal both structural and numerical matrix content while preserving determinant properties. Parallel LU decomposition is used to distribute encrypted matrix blocks across an arbitrary number of untrusted edge servers, enabling efficient and scalable determinant computation. A one-way communication model further reduces coordination overhead by eliminating inter-server interactions. To ensure result integrity with minimal client burden, we further introduce two verification algorithms: $Q_2$, a probabilistic scalar method, and $Q_3$, a deterministic and low-complexity alternative.

Mathematical analysis demonstrates that the proposed framework provides strong privacy and security guaranties, low computational overhead, and deployment flexibility—making it well-suited for secure, scalable, and real-time MDC in distributed edge-assisted systems.
\end{abstract}



\maketitle

\section{INTRODUCTION}
\IEEEPARstart{T}{he} Rapid adoption of the Internet of Things (IoT) has led to a surge in data generation and inter-device communication, necessitating scalable and energy-efficient computational frameworks. Edge computing has emerged as a viable solution, reducing latency, optimizing bandwidth, and improving response times by processing data closer to its source~\cite{Lei2015, Xiao2019}. However, its decentralized nature introduces challenges in ensuring data privacy and computational security.

Matrix determinant computation (MDC) is a fundamental operation in cryptography, machine learning, and artificial intelligence. Despite its significance, computing the determinant of large matrices remains computationally intensive. Methods such as LU decomposition incur a computational complexity of O($n^3$), making determinant calculation of large matrices impractical in resource-constrained edge environments~\cite{Xiao2019, Strang2006}. Secure outsourcing techniques, such as homomorphic encryption and secure multiparty computation (SMPC), face significant computational overhead that limits their real-time applicability~\cite{Hohenberger2005, Shan2018}. Existing parallel MDC solutions, such as the two-server model~\cite{Gao2023}, also face scalability and efficiency challenges in large-scale distributed networks.

\begin{figure*}
        \centering
        \includegraphics[width=0.87\textwidth, height=0.7\textwidth]{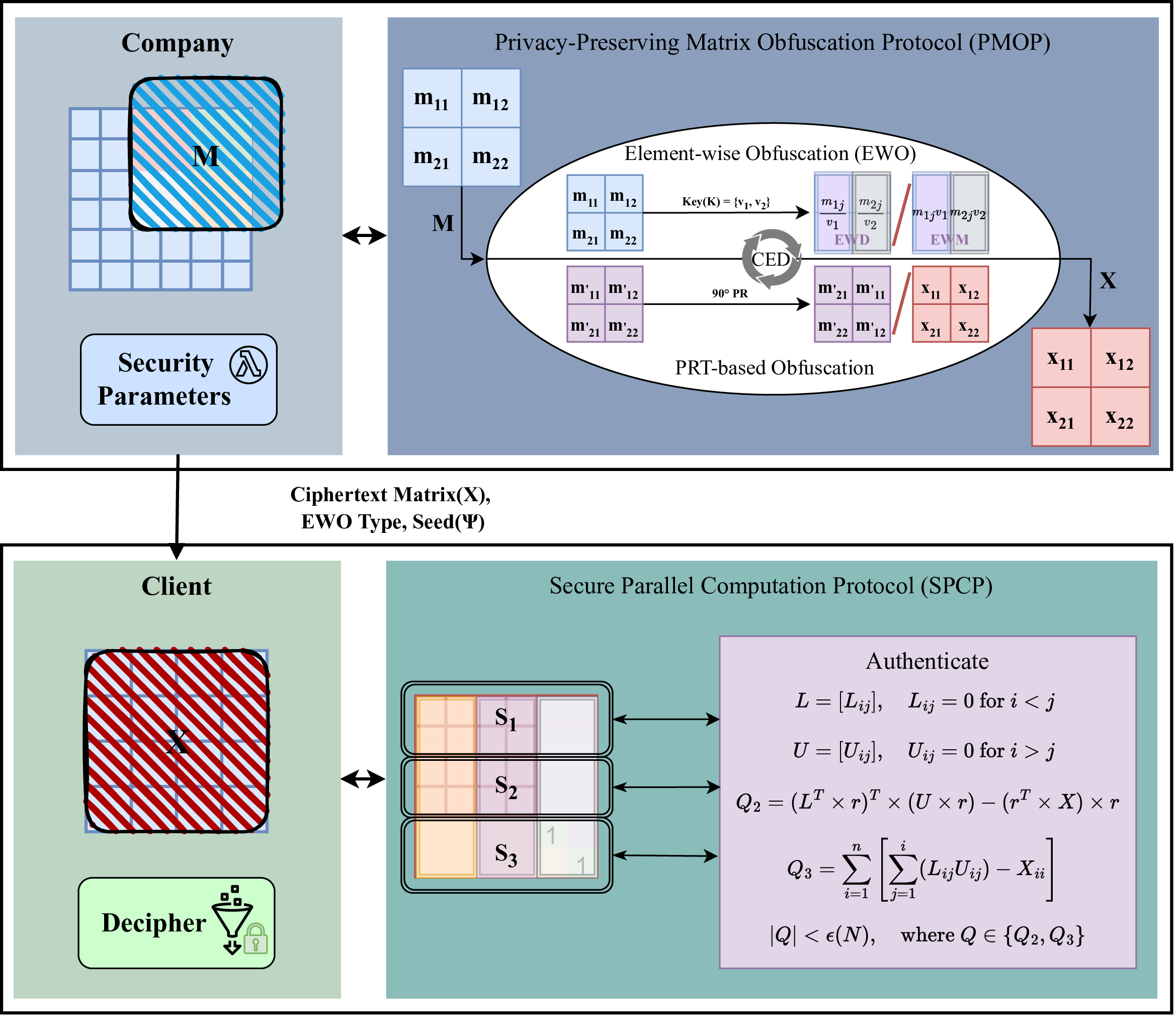} 
        \caption{Visual abstract of the framework.}
        \label{fig:1}
\end{figure*}

\begin{figure}[ht!]
    \centering
    \begin{minipage}{0.5\textwidth} 
        \centering
        \includegraphics[width=\textwidth,height=6.8cm,keepaspectratio]{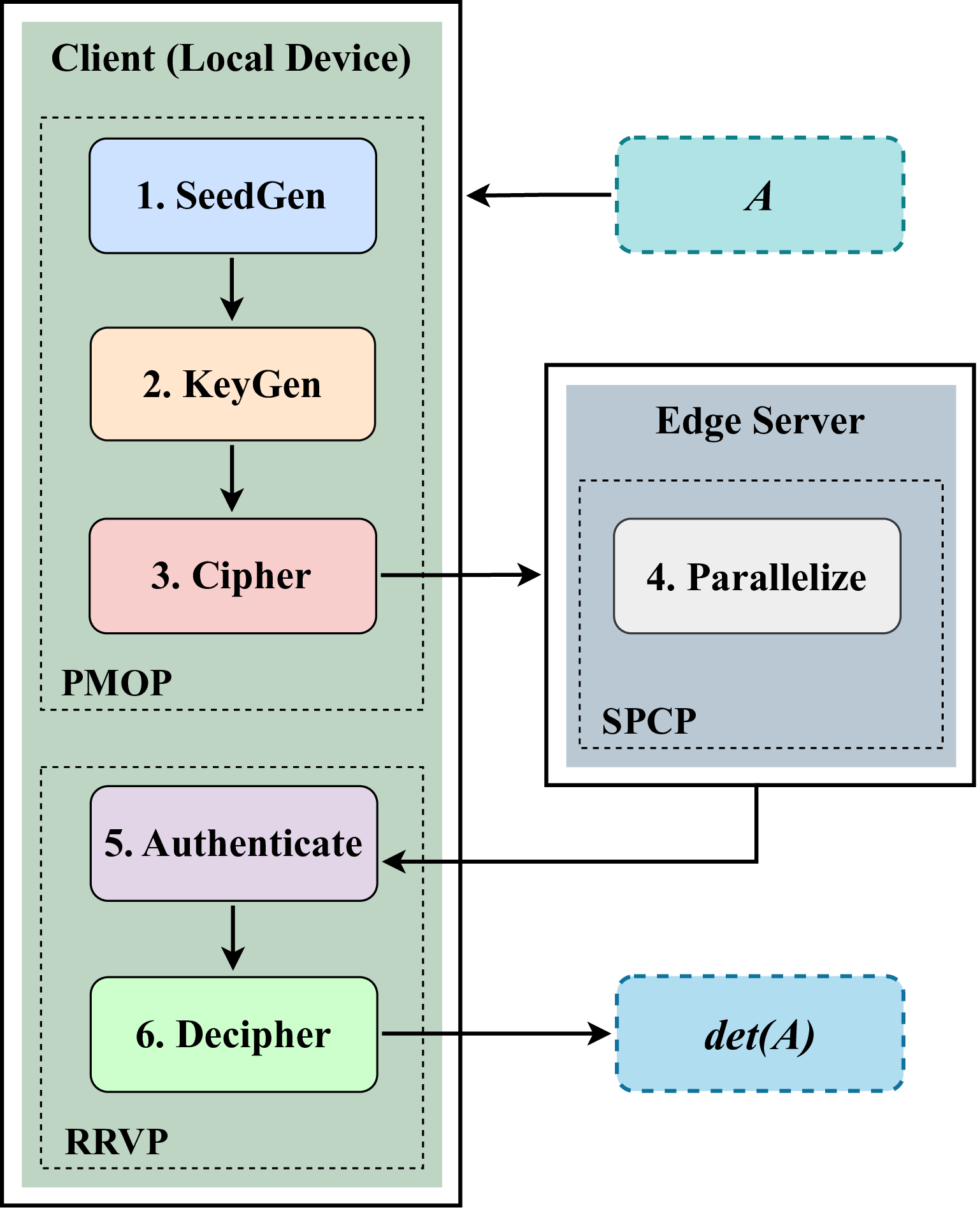}
        \vspace{-0.1mm}
    \end{minipage}
    \caption{Algorithm flow and execution placement across client and edge servers in the proposed determinant computation framework.}
    \label{fig:2}
    \vspace{-6mm}
\end{figure}

To address these limitations, this paper introduces the Secure Parallel Determinant Computation (SPDC) framework, comprising three protocols: (i) the Privacy-Preserving Matrix Obfuscation Protocol (PMOP), (ii) the Secure Parallel Computation Protocol (SPCP), and (iii) the Result Recovery and Verification Protocol (RRVP). At its core, SPDC employs the Composite Element Distortion (CED) mechanism, designed to enable efficient parallel computation of matrix determinants over an arbitrary number of untrusted edge servers. CED integrates Element-wise Obfuscation (EWO) and Panth Rotation Theorem (PRT)-based obfuscation to jointly achieve computational efficiency and strong privacy protection. Specifically, EWO secures individual matrix entries through either Element-wise Division (EWD) or Element-wise Multiplication (EWM). In contrast, PRT-based obfuscation applies a geometric rotation that alters the structural representation of the matrix without affecting its determinant, thereby introducing an additional layer of security.

Additionally, an $N$-server parallel LU decomposition model is introduced, overcoming the scalability limitations of two-server approaches and reducing computational complexity from O($n^3$) to approximately O($n^2$). The proposed model optimizes computational resource utilization, minimizes latency, and integrates a matrix augmentation technique to handle non-divisible matrix sizes while preserving determinant accuracy. Mathematical evaluations using $3$-server and $4$-server configurations demonstrate its scalability and efficiency.

To ensure result integrity, novel authentication algorithms ($Q_2$ and $Q_3$) are introduced, enabling the client to verify the correctness of results returned by edge servers with minimal computational overhead. By combining efficiency, scalability, and robust privacy mechanisms, this framework addresses critical challenges in privacy-preserving and secure MDC, making it a practical solution for real-time secure computations in distributed edge environments.

\section{CORE CONCEPTS AND TECHNIQUES EMPLOYED IN THE FRAMEWORK}

This section provides a comprehensive discussion of the mathematical techniques and computational methods used within the proposed SPDC framework. These methods are fundamental to ensuring both the efficiency of parallel processing and the confidentiality of the data during computation.

\begin{figure}[ht!]
    \centering
    \begin{minipage}{0.48\textwidth} 
        \centering
        \includegraphics[width=\textwidth,height=15cm,keepaspectratio]{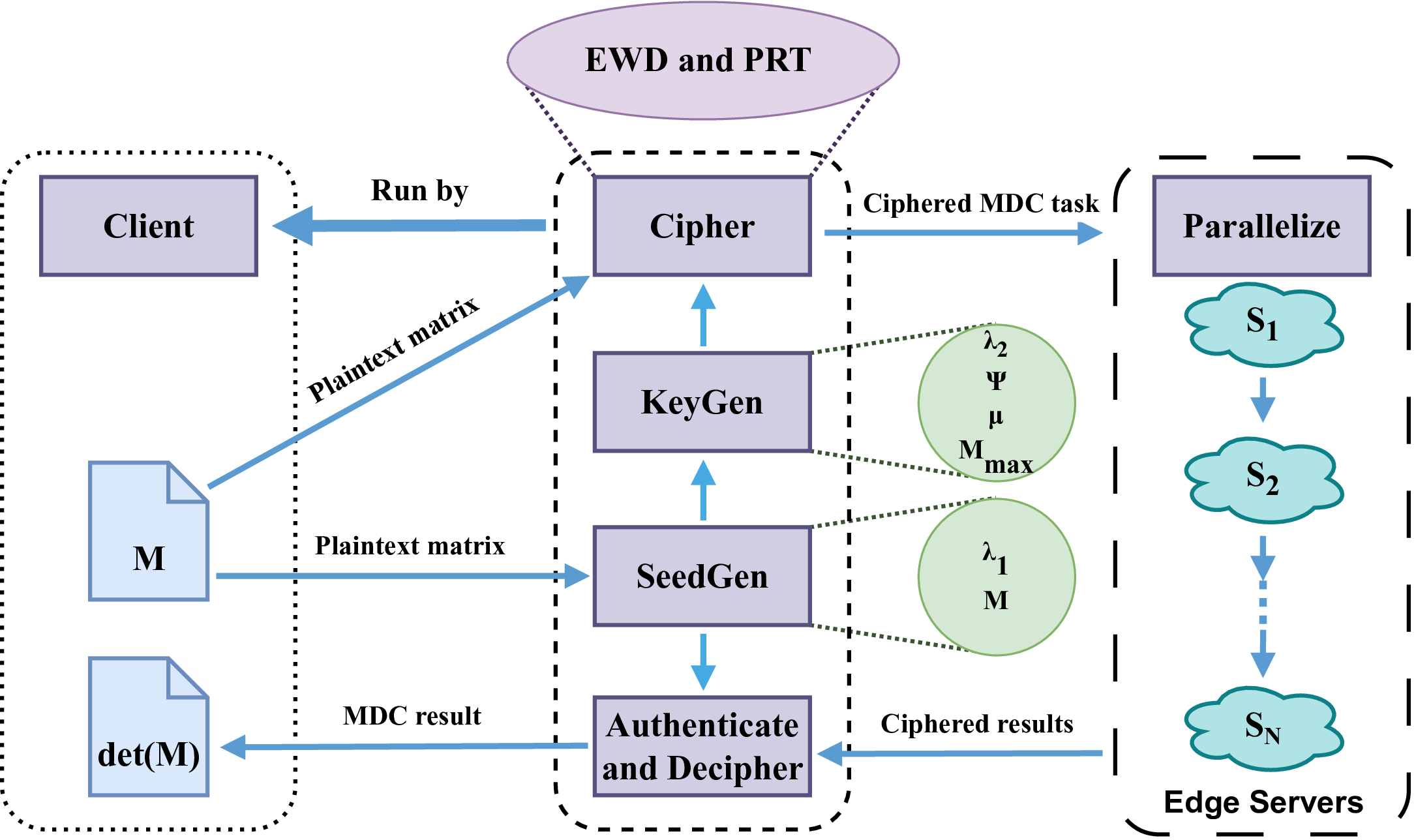}
        \vspace{-0.1mm}
    \end{minipage}
    \caption{System architecture.}
    \label{fig:3}
\end{figure}

\subsection{PANTH ROTATION THEOREM (PRT)}
The PRT is a novel contribution to matrix theory, designed to facilitate privacy-preserving and secure matrix transformations by ensuring that the absolute value of the determinant remains invariant under specific geometric transformations. This theorem plays a pivotal role in obfuscating matrix elements without affecting the accuracy of the determinant calculation.

\subsubsection{Theorem}
Let $n$ be a natural number representing the size of a square matrix. The determinant of the matrix exhibits distinct behavior under rotations of $90^\circ$, $180^\circ$, $270^\circ$, and $360^\circ$, depending on the congruence class of $n$ modulo $4$:

\subsubsubsection{For matrices of size \(n \equiv 2 \pmod{4}\) or \(n \equiv 3 \pmod{4}\):}
\vspace{0.3em}
\begin{itemize}
  \item Rotations by 90\degree{} and 270\degree{} invert the sign of the determinant.
  \item Rotations by 180\degree{} and 360\degree{} preserve the sign of the determinant.
\end{itemize}

\subsubsubsection{For matrices of size \(n \equiv 0 \pmod{4}\) or \(n \equiv 1 \pmod{4}\):}
\vspace{0.3em}
\begin{itemize}[noitemsep, topsep=5pt]
  \item Rotations by any angle (90\degree{}, 180\degree{}, 270\degree{}, and 360\degree{} do not alter the sign of the determinant.
\end{itemize}

To illustrate these transformations, consider the following $4 \times 4$ matrix $\mathbf{X}$:

\[
X =
\begin{bmatrix}
X_{11} & X_{12} & X_{13} & X_{14} \\
X_{21} & X_{22} & X_{23} & X_{24} \\
X_{31} & X_{32} & X_{33} & X_{34} \\
X_{41} & X_{42} & X_{43} & X_{44}
\end{bmatrix}
\]

The transformations of matrix $\mathbf{X}$ under various rotations are as follows:

\[
R_{90\degree{}}(X) =
\begin{bmatrix}
X_{41} & X_{31} & X_{21} & X_{11} \\
X_{42} & X_{32} & X_{22} & X_{12} \\
X_{43} & X_{33} & X_{23} & X_{13} \\
X_{44} & X_{34} & X_{24} & X_{14}
\end{bmatrix},
\]

\[
R_{180\degree{}}(X) =
\begin{bmatrix}
X_{44} & X_{43} & X_{42} & X_{41} \\
X_{34} & X_{33} & X_{32} & X_{31} \\
X_{24} & X_{23} & X_{22} & X_{21} \\
X_{14} & X_{13} & X_{12} & X_{11}
\end{bmatrix},
\]

\[
R_{270\degree{}}(X) =
\begin{bmatrix}
X_{14} & X_{24} & X_{34} & X_{44} \\
X_{13} & X_{23} & X_{33} & X_{43} \\
X_{12} & X_{22} & X_{32} & X_{42} \\
X_{11} & X_{21} & X_{31} & X_{41}
\end{bmatrix}.
\]

\subsubsection{Proof}
The proof of the PRT involves analyzing the determinant behavior under different degrees of rotation. Each case examines how column and row operations affect the determinant.

\medskip
\textsc{Case 1:} $90\degree{}$ Rotation \\
Consider an $n \times n$ matrix $\mathbf{X}$. A $90\degree{}$ clockwise rotation of $\mathbf{X}$, denoted as $R_{90\degree{}}(X)$, can be achieved by transposing $\mathbf{X}$ followed by reversing the order of the columns.

The reversal of $n$ columns can be performed through $\left[ \frac{n}{2} \right]$ swaps, where each swap inverts the determinant's sign.

Therefore,
\[
\det\big(R_{90\degree{}}(X)\big) = (-1)^{\left[ \frac{n}{2} \right]} \det(\mathbf{X})
\]

\begin{itemize}
    \item If $n = 4k$ or $n = 4k + 1$ for some positive integer $k$, then $\left[ \frac{n}{2} \right] = 2k$, which is even, leading to $\det\big(R_{90\degree{}}(X)\big) = \det(\mathbf{X})$.
    \item If $n = 4k + 2$ or $n = 4k + 3$, then $\left[ \frac{n}{2} \right] = 2k + 1$, which is odd, leading to $\det\big(R_{90\degree{}}(X)\big) = -\det(\mathbf{X})$.
\end{itemize}

\medskip
\textsc{Case 2:} $180\degree{}$ Rotation \\
A $180\degree{}$ rotation reverses the order of both rows and columns.

Since the operation involves $\left[ \frac{n}{2} \right]$ swaps for rows and $\left[ \frac{n}{2} \right]$ swaps for columns, the total number of swaps is even, leaving the determinant unchanged regardless of the matrix size.

Thus,
\[
\det\big(R_{180\degree{}}(X)\big) = (-1)^{2\left[ \frac{n}{2} \right]} \det(\mathbf{X}) = \det(\mathbf{X})
\]

\medskip
\textsc{Case 3:} $270\degree{}$ Rotation \\
A $270\degree{}$ rotation is equivalent to performing a $90\degree{}$ counterclockwise rotation, which is the same as three $90\degree{}$ clockwise rotations.

Therefore,
\[
\det\big(R_{270\degree{}}(X)\big) = (-1)^{3\left[ \frac{n}{2} \right]} \det(\mathbf{X}) = (-1)^{\left[ \frac{n}{2} \right]} \det(\mathbf{X})
\]

\begin{itemize}
    \item If $n = 4k$ or $n = 4k + 1$ for some positive integer $k$, then $\left[ \frac{n}{2} \right] = 2k$, which is even, leading to $\det\big(R_{270\degree{}}(X)\big) = \det(\mathbf{X})$.
    \item If $n = 4k + 2$ or $n = 4k + 3$, then $\left[ \frac{n}{2} \right] = 2k + 1$, which is odd, leading to $\det\big(R_{270\degree{}}(X)\big) = -\det(\mathbf{X})$.
\end{itemize}

\medskip
\textsc{Case 4:} $360\degree{}$ Rotation \\
A $360\degree{}$ rotation returns the matrix to its original configuration:
\[
\det\big(R_{360\degree{}}(X)\big) = (-1)^{4\left[ \frac{n}{2} \right]} \det(\mathbf{X}) = \det(\mathbf{X})
\]
Thus for any $n$, $\det\!\big(R_{360^\circ}(X)\big)$ does not invert the sign.

\subsubsection{Implications}
The PRT plays a pivotal role in enabling privacy-preserving matrix transformations. It guarantees that the absolute value of a determinant remains invariant under specific rotations, thereby safeguarding sensitive matrix data during outsourced computations. By bridging geometric transformations with algebraic properties, PRT contributes simultaneously to applied cryptography and theoretical linear algebra. In image processing, where matrices represent pixel data, PRT enables fast and secure rotations ($90\degree{}$, $180\degree{}$, $270\degree{}$) without altering fundamental image characteristics, thus supporting image encryption. Overall, PRT prevents information leakage during computation, ensures accurate determinant recovery after decryption, and strengthens the design of secure data outsourcing protocols.

\subsection{MATRIX AUGMENTATION}
In distributed systems, matrix augmentation ensures efficient parallel processing when the matrix size is not divisible by the number of servers. It allows balanced load distribution while preserving the determinant.

\subsubsection{Definition and Statement}
Let $A$ be an $n \times n$ square matrix. Consider an augmented $(n+1) \times (n+1)$ matrix $B$, where:
\begin{itemize}
    \item The upper-left $n \times n$ submatrix of $B$ is identical to $A$,
    \item The element at position $(n+1,\, n+1)$ is set to $1$,
    \item Either the remaining elements of the $(n+1)$-th row or the $(n+1)$-th column are filled with zeros, and
    \item The remaining $n$ elements (in either the row or column not filled with zeros) may contain any real numbers.
\end{itemize}
Under these conditions, $\det(B) = \det(A)$.

\subsubsection{Relation to Block Triangular Matrices}
The augmentation process in this framework can be viewed as a specific instance of block matrices, where additional rows and columns are introduced without altering the determinant. The augmented matrix $B$ can be expressed as:
\[
B = \begin{pmatrix}
A & 0 \\
R & C
\end{pmatrix},
\]
where $A$ is the original $n \times n$ matrix, $R$ is a row vector of size $1 \times n$ containing real numbers, and $C$ is a scalar set to one. The addition of $R$ and a zero-filled column ensures that the determinant of the augmented matrix remains unchanged from the original matrix.

\subsubsection{Implication}
\begin{itemize}[noitemsep, topsep=5pt]
    \item Facilitates parallel computation by ensuring equally sized submatrices for all servers.
    \item Allows even-sized partitioning, making it compatible with scalable parallel LU decomposition.
    \item Improves computational efficiency in distributed systems by optimizing task distribution while maintaining determinant accuracy.
\end{itemize}

\subsection{LU FACTORIZATION}
LU factorization decomposes a nonsingular $n \times n$ matrix $A$ into a lower triangular matrix $L$ and an upper triangular matrix $U$, expressed as $A = L \times U$ \cite{Strang2006}. This simplifies determinant computation since the determinant of a triangular matrix is the product of its diagonal elements. In distributed systems, parallelizing LU factorization improves efficiency by distributing workloads for faster computation.

\section{DEFINITIONS}
In this section, we introduce the SPDC framework and define its associated security guarantees.

\begin{figure}[ht!]
    \centering
    \begin{minipage}{0.48\textwidth} 
        \centering
        \includegraphics[width=\textwidth,height=2cm]{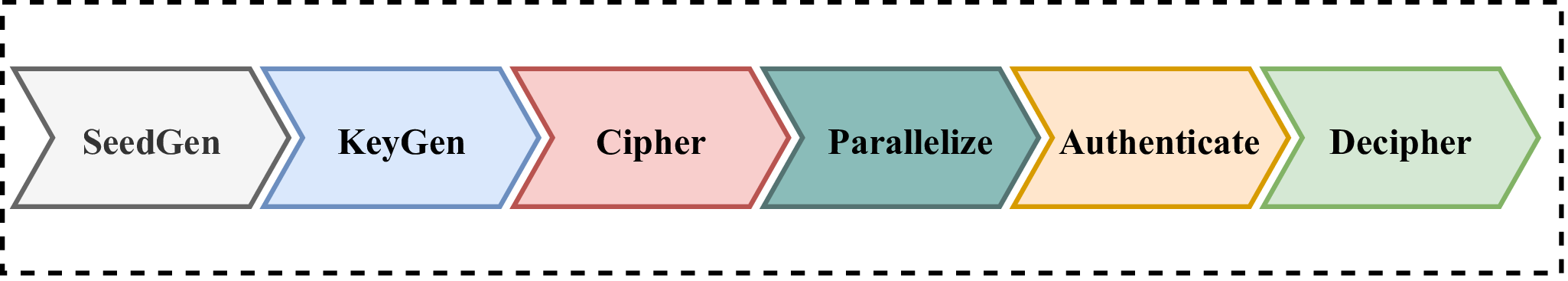}
        \vspace{-0.2mm}
    \end{minipage}
    \caption{Process workflow.}
    \label{fig:4}
\end{figure}

\subsection{FRAMEWORK}
Our SPDC framework consists of three protocols (PMOP, SPCP, and RRVP), each implemented using a set of algorithms. The framework is represented as a tuple of six core algorithms, (\allowbreak\text{SeedGen}, \allowbreak\text{KeyGen}, \allowbreak\text{Cipher}, \allowbreak\text{Parallelize}, \allowbreak\text{Authenticate}, \allowbreak\text{Decipher}), where each algorithm is formalized as follows:

\begin{enumerate}
    \item {\raggedright\textbf{Seed Generation Algorithm:}\par}
    $\text{SeedGen}(\lambda_{1}, \mathbf{M}) \rightarrow (\Psi, \mu, M_{\max})$.
    Given the input matrix $\mathbf{M}$ and security parameter $\lambda_{1}$, the client generates a secure pseudo-random number, referred to as $\text{Seed}(\Psi)$, based on the matrix's statistical characteristics (mean $\mu$ and maximum value $M_{\max}$) along with $\lambda_{1}$. This seed is used for subsequent key generation and encryption.
    
    \vspace{0.5em}
    \item {\raggedright\textbf{Key Generation Algorithm:}\par} $\text{KeyGen}(\lambda_{2}, \Psi, \mu, M_{\max}) \rightarrow K$. 
    With the seed $\Psi$, matrix properties $\mu$ and $M_{\max}$, and an additional security parameter $\lambda_{2}$, the client generates a secret key $K$, which includes a blinding vector $\mathbf{v}$ to obfuscate the matrix during encryption.
    
    \vspace{0.5em}
    \item {\raggedright\textbf{Data Protection Algorithm:}\par}
    $\text{Cipher}(K, \mathbf{M}) \rightarrow \mathbf{X}$.  
    The client uses the secret key $K$ to encrypt the matrix $\mathbf{M}$, resulting in the ciphertext matrix $\mathbf{X}$. The encryption process involves EWD by the blinding vector $\mathbf{v}$ and the application of the PRT to maintain the matrix's determinant properties while obfuscating its structure.
    
    \vspace{0.5em}
    \item {\raggedright\textbf{Parallel Computation Algorithm:}\par}
    $\text{Parallelize}(\mathbf{X}) \rightarrow (L, U)$.  
    For the ciphertext matrix $\mathbf{X}$, edge servers compute the LU decomposition in parallel. The matrix is divided into submatrices and distributed to an arbitrary number of untrusted servers for efficient computation of the matrix determinant.

    \vspace{0.5em}
    \item {\raggedright\textbf{Result Authentication Algorithm:}\par}
    $\text{Authenticate}(L, U, \mathbf{X}) \rightarrow \{1,0\}$.  
    After receiving the LU decomposition results from the servers, the client verifies their correctness. 
    If the results are valid, the output is $1$; otherwise, it is $0$. 
    Authentication is performed using lightweight authentication formulas $Q_{2}$ and $Q_{3}$, which reduce computational load.

    \vspace{0.5em}
    \item {\raggedright\textbf{Result Decryption Algorithm:}\par}
    $\text{Decipher}(\Psi, L, U) \rightarrow \det(\mathbf{M})$.  
    The client uses the seed $\Psi$ and the verified LU decomposition results to compute the determinant of the original matrix $\mathbf{M}$, ensuring correctness and privacy of the outsourced computation.

\end{enumerate}

\noindent
\subsection{SECURITY DEFINITION}
A SPDC framework must ensure the security of both the matrix and its determinant, even when outsourced to untrusted edge servers. This guarantees data confidentiality, computational integrity, and protection against adversarial attacks.
\medskip

\textit{Definition 1 (Blinding Vector and Rotational Obfuscation Security).}
The matrix encryption process employs a blinding vector derived from the matrix seed $\Psi$ in combination with rotational obfuscation via the Panth Rotation Theorem (PRT), thereby concealing both the matrix structure and its determinant.

Given an encrypted matrix $\mathbf{X}$, an adversary's probability $\Pr$ of inferring the original matrix $\mathbf{M}$ or its determinant is negligible, bounded by $\epsilon(\lambda)$, where $\lambda$ is the security parameter.

\begin{itemize}
    \item \textbf{Indistinguishability:} For any adversary $\mathcal{A}$ attempting to infer the original matrix $\mathbf{M}$ or its determinant from the encrypted matrix $\mathbf{X}$, the probability $\Pr$ of success is negligible. Formally, 
    \[
        \Pr[\mathcal{A}(\mathbf{X}) \to \mathbf{M} \text{ or } \det(\mathbf{M})] \leq \epsilon(\lambda),
    \]
    where $\epsilon(\lambda)$ is a negligible function in $\lambda$. As the security parameter $\lambda$ increases, the probability of a successful attack approaches zero.
    \item \textbf{Determinant Preservation:} The determinant of the encrypted matrix differs from that of the original matrix but can be fully restored upon decryption with the seed $\Psi$, following the relationship:
    \[
        \det(\mathbf{M}) = \det(A) \times \text{Correction Factor}
    \]
\end{itemize}

\textit{Definition 2 (Correctness of Outsourced Computation).}
 After parallel LU decomposition is performed on the encrypted matrix $\mathbf{X}$, the client must verify the correctness of the results \cite{Gao2022}. This is done through lightweight authentication algorithms that confirm the validity of the reconstructed matrix $X' = L \times U$, where $L$ and $U$ are the LU components returned by edge servers.

\begin{itemize}
    \item \textbf{Consistency:} The reconstructed matrix $X'$ approximates the original encrypted matrix $\mathbf{X}$ within a small threshold $\epsilon$, ensuring that deviations are only due to numerical inconsistencies, not malicious tampering:
    \[
    |\det(X') - \det(\mathbf{X})| < \epsilon
    \]
    \item \textbf{Efficient Authenticity:} The authentication process relies on lightweight computations, ensuring that the client, even with limited computational resources, can efficiently verify the correctness of the LU decomposition computation.
\end{itemize}

\vspace{0.7em}
\textit{Definition 3 (Resilience to Collusion).}
This security property of the framework ensures resilience to collusion between multiple edge servers. Even if a subset of servers colludes to share intermediate results, they cannot reconstruct the original matrix or its determinant. This is achieved through the distribution of encrypted submatrices and the sequential one-way communication pattern between servers.

The probability $\Pr$ that a subset of servers $S'$ can successfully reconstruct the matrix $\mathbf{M}$ or its determinant from the encrypted data $\mathbf{X}$ is negligible, expressed mathematically as:
\[
\Pr[S'(\mathbf{X}) \to \mathbf{M} \text{ or } \det(\mathbf{M})] < \epsilon(\lambda)
\]

This definition ensures resilience against potential collusion among servers, thereby preserving the confidentiality of the matrix across distributed environments. By preventing any individual or group of servers from gaining insights into the data, this framework maintains data confidentiality and robustness in multi-server architectures.

\section{PROTOCOL CONSTRUCTION}
We construct a SPDC framework for MDC, consisting of six key stages: \emph{Seed\-Gen}, \emph{Key\-Gen}, \emph{Cipher}, \emph{Parallel\-ize}, \emph{Authen\-ticate}, and \emph{Decipher}. 
This framework enables secure outsourcing of MDC tasks to an arbitrary number of untrusted edge servers, ensuring both data confidentiality and computational integrity through advanced cryptographic and algebraic methods. Each stage is carefully designed to maintain privacy, optimize efficiency, and ensure an accurate recovery of the matrix determinant.

\subsection{SEED GENERATION ALGORITHM (SEEDGEN)}
\noindent\textbf{SeedGen}$(\lambda_1, \mathbf{M}) \rightarrow (\Psi, \mu, M_{\max})$. The SeedGen algorithm generates a secure seed $\Psi$ based on the input matrix $\mathbf{M} = (m_{ij})$ and cryptographic security parameter $\lambda_1$.

To generate $\Psi$, the algorithm first computes the statistical properties of the matrix, specifically the mean $\mu$ and the maximum value $M_{\max}$, defined as:

\[
\mu = \frac{1}{n^2} \sum_{i=1}^n \sum_{j=1}^n m_{ij}
\]
\vspace{-0.5em}
\noindent
\[
M_{\max} = \max(m_{ij}), \quad \forall i,j
\]

These statistical properties provide matrix-specific randomness, ensuring that the seed reflects the internal structure of the matrix. The security parameter $\lambda_1$, combined with these statistical properties, contributes to the cryptographic strength of the seed. The seed $\Psi$ is computed as:
\vspace{-0.2em}
\noindent
\[
\Psi = \mathbf{H}(\lambda_1, \mu, M_{\max})
\]

where $\mathbf{H}$ is a cryptographic hash function, ensuring that $\Psi$ is unpredictable and resistant to adversarial attacks. The computed seed $\Psi$, along with the matrix's mean $\mu$ and maximum $M_{\max}$, will be used in the key generation process to ensure secure randomization and matrix obfuscation.

\subsection{KEY GENERATION ALGORITHM (KEYGEN)}
\noindent\textbf{KeyGen}$(\lambda_2, \Psi, \mu, M_{\max}) \rightarrow K$. 
The KeyGen algorithm generates the secret key $K$ used for encryption. 
It relies on the seed $\Psi$, matrix properties $\mu$ and $M_{\max}$, 
and the cryptographic parameter $\lambda_2$.

\medskip
\noindent
Given the seed $\Psi$, the matrix properties, the key generation algorithm constructs a blinding vector 
$\mathbf{v} = [v_1, v_2, \dots, v_n]$, which is used to obfuscate the matrix elements during encryption. 
The elements of the blinding vector $\mathbf{v}$ are selected to satisfy the condition:
\[
\prod_{i=1}^n v_i = \Psi, \quad v_i \neq 1 \ \forall i \in [1,n]
\]

\noindent
The blinding vector $\mathbf{v}$ is generated using a cryptographically secure pseudorandom number generator (CSPRNG or any), which guarantees the security and unpredictability of the vector elements. This ensures that even if an adversary obtains the ciphertext matrix $\mathbf{X}$, they will be unable to reconstruct the original matrix $\mathbf{M}$ without access to the complete blinding vector. The secret key $K = \{\mathbf{v}\}$ is securely stored by the client and is used to cipher the matrix $\mathbf{M}$ into its encrypted form $\mathbf{X}$.

\subsection{DATA PROTECTION ALGORITHM (CIPHER)}
\noindent\textbf{Cipher}$(K, \mathbf{M}) \rightarrow \mathbf{X}$. 
The Cipher algorithm applies CED, a dual-layer encryption method combining EWO and PRT-based Obfuscation to secure both matrix values and structure efficiently. These obfuscation techniques run simultaneously, not only to enhance security but also to reduce the time complexity of ciphering.

\subsubsection{Element-wise Obfuscation (EWO):}
It obfuscates each element of matrix $\mathbf{M}$ through one of two distinct transformations:

\subsubsubsection{Element-wise Division (EWD):}
Each element $m_{ij}$ of the original matrix is divided by the corresponding element of the blinding vector:
\[
m'_{ij} = \frac{m_{ij}}{v_i}
\]

\subsubsubsection{Element-wise Multiplication (EWM):}
Alternatively, the matrix can be transformed by multiplying each element $m_{ij}$ of $\mathbf{M}$ 
by the corresponding element of the blinding vector:
\[
m'_{ij} = m_{ij} \times v_i
\]
These transformations conceal individual values, preventing unauthorized access.

\subsubsection{PRT-based Obfuscation:}
In addition to EWO, the matrix simultaneously undergoes PRT-based Obfuscation, a rotational obfuscation technique based on PRT. This obfuscation preserves the matrix determinant while altering its spatial arrangement by rotating the matrix by a fixed angle (\(90^\circ\), \(180^\circ\), or \(270^\circ\)), depending on the seed value \(\Psi\). The degree of rotation is determined by the \(\textit{Rotate}(\Psi)\) function:
\vspace{-0.2em}
\[
\text{Rotate}(\Psi) =
\begin{cases}
1 & \text{if } 90^\circ \text{ rotation is required} \\
2 & \text{if } 180^\circ \text{ rotation is required} \\
3 & \text{if } 270^\circ \text{ rotation is required}
\end{cases}
\]

Given a seed value \(\Psi\), the \(\textit{Rotate}(\Psi)\) function determines the appropriate rotation, returning a value in the set \(\{1, 2, 3\}\) that corresponds to the rotation angles. The seed value \(\Psi\) may initially be a floating-point number. To convert \(\Psi\) into a Quantized Seed (\(\Psi'\)) for rotation selection, one of the following methods is employed:
\begin{itemize}
    \item Floor Function: \(\Psi' = \lfloor \Psi \rfloor\)
    \item Ceiling Function: \(\Psi' = \lceil \Psi \rceil\)
    \item Round Function: \(\Psi' = \text{round}(\Psi)\)
    \item Truncation: \(\Psi' = \text{int}(\Psi)\)
\end{itemize}

Once \(\Psi'\) is derived, it is mapped to the rotation set \(\{1, 2, 3\}\) using various methods, one of which is illustrated below:
\[
\text{Rotation Degree} = (\Psi' \bmod 3) + 1
\]

This ensures a pseudo-random yet deterministic rotation based on the seed value, making it infeasible for an adversary to reconstruct the original matrix even if some element values are known.

\subsection{PARALLEL COMPUTATION ALGORITHM (PARALLELIZE)}
\noindent \textbf{Parallelize}$(\mathbf{X}) \rightarrow (L,U)$. The Parallelize algorithm enhances the efficiency of MDC by distributing workloads across an arbitrary number of untrusted edge servers, overcoming the scalability limits of traditional $2$-server MDC frameworks.

Unlike recent work such as \cite{Gao2023}, which implements parallel determinant outsourcing using a fixed $2$-server architecture ($N = 2$), our protocol introduces a generalized formulation that supports any number of servers $N \geq 2$. The underlying structure is formally parameterized, enabling deployment across diverse edge environments with varying numbers of computational nodes. When instantiated with $N=2$, our formulation reduces to the approach in \cite{Gao2023}, thereby preserving compatibility while extending parallelism to arbitrary configurations.

Additionally, for input matrices whose dimensions are not evenly divisible by $N$, a determinant-preserving augmentation technique is applied. This ensures uniform partitioning across all servers without compromising correctness, thereby maintaining computational balance and accuracy under varied input scenarios.

\subsubsection{Matrix Partitioning and Augmentation}
The client $C$ partitions the encrypted input matrix $\mathbf{X}$ into smaller submatrices and distributes them among edge servers for parallel processing. If the input matrix of size $n \times n$ is not evenly divisible by the number of servers $N$ such that the resulting submatrix size is greater than $1$, the matrix is augmented. Specifically, the matrix is augmented to a size of $(n+p) \times (n+p)$, where $p$ is the minimum positive integer that ensures both:

\begin{itemize}
    \item $\frac{n+p}{N}$ is an integer, and
    \item $\frac{n+p}{N} > 1$.
\end{itemize}

After augmentation, the matrix is partitioned into $N^2$ submatrices, each of size $\left( \frac{n+p}{N} \right) \times \left( \frac{n+p}{N} \right)$.

\medskip
\subsubsubsection{\textbf{Examples}}
\textbf{Example 1: 3-Server Setup with a $4 \times 4$ Input Matrix} \\
For $N=3$ and an input matrix of size $4 \times 4$, we require
\[
\frac{4+p}{3} > 1 \quad \text{and} \quad \frac{4+p}{3} \in \mathbb{Z}.
\]
The smallest $p$ that satisfies these conditions is $p=2$, so the augmented matrix becomes $(4+2) \times (4+2) = 6 \times 6$.  
This matrix is then partitioned into $3^2 = 9$ submatrices, each of size $(3/2) \times (3/2) = 2 \times 2$.

\medskip
\noindent\textbf{Example 2: 2-Server Setup with a $6 \times 6$ Input Matrix} \\
For $N=2$ and an input matrix of size $6 \times 6$, we compute 
\[
\frac{6+p}{2}.
\]

Since $6$ is already divisible by $2$ (with $\frac{6}{2} = 3 > 1$), we set $p=0$.  
The matrix is partitioned into $2^2 = 4$ submatrices, each of size $3 \times 3$.  
In this configuration, each server is assigned two submatrices.

\medskip
\subsubsubsection{\textbf{Assignment of Submatrices}}
In all configurations, the $N^2$ submatrices are allocated among the $N$ servers according to a row-wise scheme. For example, in the generalized $N$-server setup, if the submatrices are labeled $X_{ij}$ (where $i, j = 1, 2, \ldots, N$), then:
\vspace{0.2em}
\begin{itemize}
    \item \textbf{Server $S_1$:} Handles submatrices \\
    \centerline{$X_{11}, X_{12}, X_{13}, \ldots, X_{1N}$}
    
    \item \textbf{Server $S_2$:} Handles submatrices \\
    \centerline{$X_{21}, X_{22}, X_{23}, \ldots, X_{2N}$}
    
    \item \textbf{Server $S_3$:} Handles submatrices \\
    \centerline{$X_{31}, X_{32}, X_{33}, \ldots, X_{3N}$}
    
    \item[\phantom{$\bullet$}] \centerline{$\vdots$}
    \vspace{-1em}
    \item \textbf{Server $S_N$:} Handles submatrices \\
    \centerline{$X_{N1}, X_{N2}, X_{N3}, \ldots, X_{NN}$}
\end{itemize}
\vspace{0.2em}

Following the initial LU decomposition on their respective submatrices, servers exchange the necessary intermediate results to the next server to ensure the overall computational integrity.

\subsubsection{The Algorithm}
The parallel LU decomposition model for two servers has been previously established in the literature \cite{Gao2023}. We extend this by implementing a scalable model across multiple edge servers, including $3$-server, $4$-server, and generalized $N$-server configurations. Starting with the $3$-server setup, we demonstrate improved efficiency over the $2$-server model. This foundation extends to $4$ servers, illustrating scalability for complex computations, and ultimately generalizes to $N$-servers for efficiently handling larger datasets and utilizing greater computational resources.

\vspace{-1mm}
\begin{figure}[ht!]
    \centering
    \begin{minipage}[]{0.49\textwidth} 
        \centering
        \includegraphics[width=\textwidth, height=10cm]{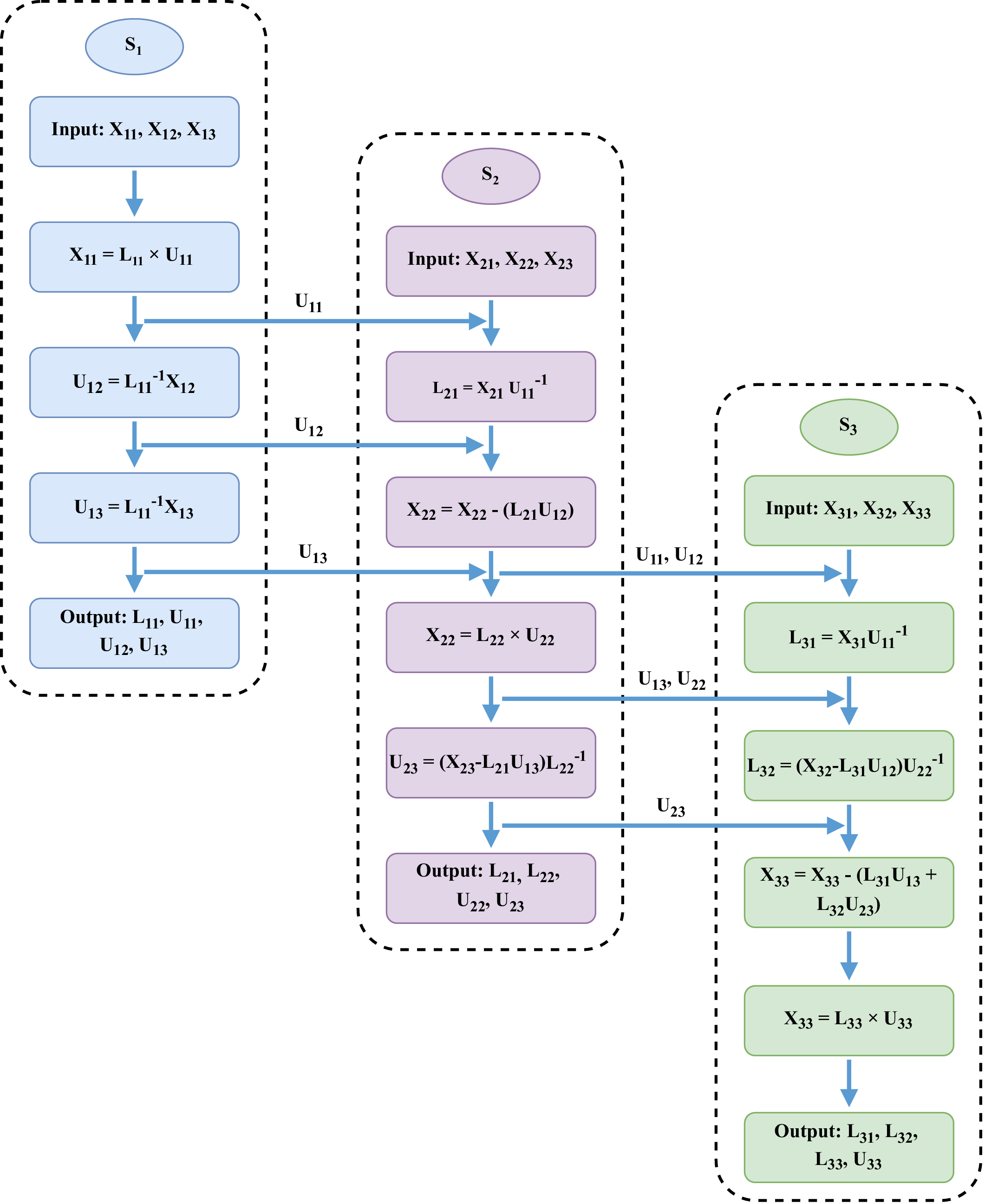}
        \caption{Three-server parallel LU decomposition process.}
        \label{fig:5}
    \end{minipage}\hfill
\end{figure}
\vspace{-2mm}

\subsubsubsection{Three-Server LU Decomposition Algorithm}
\medskip
\algheader{3-Server Parallel Computing}

\textbf{Input:} The ciphertext blocks $X_{ij}$ ($1 \le i,j \le 3$).\\
\textbf{Output:} The computation results $L$ and $U$. \\

\textbf{Initial Phase:}
\step{1}{ S$_1$ performs LU factorization on $X_{11}$ s.t. $X_{11}=L_{11}\times U_{11}$ and sends $U_{11}$ to S$_2$.}
\medskip
\vspace{1mm}
\textbf{First Computation \& Communication Phase:}
\step{2}{ S$_1$ computes $U_{12}=L_{11}^{-1}X_{12}$, then sends $U_{12}$ to S$_2$.}
\step{3}{ S$_2$ activates and computes $L_{21}=X_{21}U_{11}^{-1}$.}
\medskip

\textbf{Second Computation \& Communication Phase:}
\step{4}{ S$_1$ computes $U_{13}=L_{11}^{-1}X_{13}$ and sends $U_{13}$ to S$_2$.}
\step{5}{ S$_2$ updates $X_{22}=X_{22}-(L_{21}U_{12})$.}
\step{6}{ S$_2$ sends $\{U_{11},U_{12}\}$ to S$_3$.}
\medskip

\textbf{Third Computation \& Communication Phase:}
\step{7}{ S$_1$ returns result $\mathrm{res}_1=\{L_{11},U_{11},U_{12},U_{13}\}$ to C.}
\step{8}{ S$_2$ performs $X_{22}=L_{22}\times U_{22}$.}
\step{9}{ S$_2$ sends $\{U_{13},U_{22}\}$ to S$_3$.}
\step{10}{ S$_3$ activates and computes $L_{31}=X_{31}U_{11}^{-1}$.}
\medskip

\textbf{Fourth Computation \& Communication Phase:}
\step{11}{ S$_2$ computes $U_{23}=L_{22}^{-1}(X_{23}-L_{21}U_{13})$.}
\step{12}{ S$_2$ sends $U_{23}$ to S$_3$.}
\step{13}{ S$_3$ computes $L_{32}=U_{22}^{-1}(X_{32}-L_{31}U_{12})$.}
\medskip

\textbf{Final Computation Phase:}
\step{14}{ S$_2$ returns result $\mathrm{res}_2=\{L_{21},L_{22},U_{22},U_{23}\}$ to C.}
\step{15}{ S$_3$ computes $X_{33}=X_{33}-(L_{31}U_{13}+L_{32}U_{23})$, \\then $X_{33}=L_{33}\times U_{33}$.}
\medskip
\step{16}{ S$_3$ returns result $\mathrm{res}_3=\{L_{31},L_{32},L_{33},U_{33}\}$ to C.}
\medskip

\textbf{Final Output Integration:}
\step{17}{ C integrates $\mathrm{res}_1,\mathrm{res}_2,$ and $\mathrm{res}_3$ into $L$ and $U$.}

\vspace{6pt}\hrule height 0.8pt
\medskip

\subsubsubsection{Four-Server LU Decomposition Algorithm}
\medskip
\algheader{4-Server Parallel Computing}

\textbf{Input:} The ciphertext blocks $X_{ij}$ ($1 \le i,j \le 4$). \\
\textbf{Output:} The computation results $L$ and $U$. \\

\textbf{Initial Phase:}
\step{1}{S$_1$ performs LU factorization on $X_{11}$ s.t. $X_{11}=L_{11}\times U_{11}$ and sends $U_{11}$ to S$_2$.}
\medskip

\textbf{First Computation \& Communication Phase:}
\step{2}{S$_1$ computes $U_{12}=L_{11}^{-1}X_{12}$, then sends $U_{12}$ to S$_2$.}
\step{3}{S$_2$ activates and computes $L_{21}=X_{21}U_{11}^{-1}$.}
\medskip

\textbf{Second Computation \& Communication Phase:}
\step{4}{S$_1$ computes $U_{13}=L_{11}^{-1}X_{13}$ and sends $U_{13}$ to S$_2$.}
\step{5}{S$_2$ updates $X_{22}=X_{22}-(L_{21}U_{12})$.}
\step{6}{S$_2$ sends $\{U_{11},U_{12}\}$ to S$_3$.}
\medskip

\textbf{Third Computation \& Communication Phase:}
\step{7}{S$_1$ computes $U_{14}=L_{11}^{-1}X_{14}$.}
\step{8}{S$_1$ sends $U_{14}$ to S$_2$.}
\step{9}{S$_2$ performs $X_{22}=L_{22}\times U_{22}$.}
\step{10}{S$_2$ sends $\{U_{13},U_{22}\}$ to S$_3$.}
\step{11}{S$_3$ activates and computes $L_{31}=X_{31}U_{11}^{-1}$.}
\medskip

\textbf{Fourth Computation \& Communication Phase:}
\step{12}{S$_1$ returns result $\mathrm{res}_1=\{L_{11},U_{11},U_{12},U_{13},U_{14}\}$ \\to C.}
\step{13}{S$_2$ computes $U_{23}=(X_{23}-L_{21}U_{13})L_{22}^{-1}$.}
\step{14}{S$_2$ sends $\{U_{14},U_{23}\}$ to S$_3$.}
\step{15}{S$_3$ computes $L_{32}=(X_{32}-L_{31}U_{12})U_{22}^{-1}$.}
\step{16}{S$_3$ sends $\{U_{11},U_{12},U_{13},U_{22}\}$ to S$_4$.}
\medskip

\textbf{Fifth Computation \& Communication Phase:}
\step{17}{S$_2$ computes $U_{24}=(X_{24}-L_{21}U_{14})L_{22}^{-1}$.}
\step{18}{S$_2$ sends $U_{24}$ to S$_3$.}
\step{19}{S$_3$ updates $X_{33}=X_{33}-(L_{31}U_{13}+L_{32}U_{23})$.}
\step{20}{S$_4$ activates and computes $L_{41}=X_{41}U_{11}^{-1}$.}
\medskip

\textbf{Sixth Computation \& Communication Phase:}
\step{21}{S$_2$ returns result $\mathrm{res}_2=\{L_{21},L_{22},U_{22},U_{23},U_{24}\}$ \\to C.}
\medskip
\step{22}{S$_3$ performs $X_{33}=L_{33}\times U_{33}$.}
\step{23}{S$_3$ sends $\{U_{24},U_{33}\}$ to S$_4$.}
\step{24}{S$_4$ computes $L_{42}=(X_{42}-L_{41}U_{12})U_{22}^{-1}$.}
\medskip

\textbf{Seventh Computation \& Communication Phase:}
\step{25}{S$_3$ computes $U_{34}=(X_{34}-(L_{31}U_{14}+L_{32}U_{24}))L_{33}^{-1}$.}
\step{26}{S$_3$ sends $U_{34}$ to S$_4$.}
\step{27}{S$_4$ computes $L_{43}=(X_{43}-(L_{41}U_{13}+L_{42}U_{23}))U_{33}^{-1}$.}
\medskip

\textbf{Eighth Computation \& Communication Phase:}
\step{28}{S$_3$ returns result $\mathrm{res}_3=\{L_{31},L_{32},L_{33},U_{33},U_{34}\}$ \\to C.}
\medskip
\step{29}{S$_4$ updates $X_{44}=X_{44}-(L_{41}U_{14}+L_{42}U_{24}+L_{43}U_{34})$.}
\step{30}{S$_4$ performs $X_{44}=L_{44}\times U_{44}$.}
\step{31}{S$_4$ returns result $\mathrm{res}_4=\{L_{41},L_{42},L_{43},L_{44},U_{44}\}$ \\to C.}
\medskip

\textbf{Final Output Integration:}
\step{32}{Client C integrates $\mathrm{res}_1,\mathrm{res}_2,\mathrm{res}_3,\mathrm{res}_4$ into $L$ and $U$.}

\vspace{6pt}\hrule height 0.8pt

\medskip
\subsubsubsection{N-Server LU Decomposition Algorithm}
\medskip
\algheader{N-Server Parallel Computing}

\textbf{Input:} The ciphertext blocks $X_{ij}$ ($1 \le i,j \le N$). \\
\textbf{Output:} The computation results $L$ and $U$. \\

\textbf{$S_1$ Computation:}
\step{1}{Performs LU factorization on $X_{11}$ s.t.\\ 
$X_{11} = L_{11} \times U_{11}$.}
\step{2}{Computes $U_{1j} = L_{11}^{-1} X_{1j}$ (for $j = 2,\dots,N$).}
\begin{itemize}[leftmargin=*, nosep]
  \item Sends $U_{1j}$ (for $j = 1,\dots,N$) to $S_2$.
  \item Returns result $\mathrm{res}_1 = \{L_{11}, U_{1j}\}$ ($j = 1,\dots,N$) to $C$.
\end{itemize}
\medskip

\textbf{$S_2$ Computation:}
\step{3}{Computes $L_{21} = X_{21} U_{11}^{-1}$.}
\step{4}{Updates $X_{22} = X_{22} - (L_{21} U_{12})$.}
\step{5}{Performs $X_{22} = L_{22} \times U_{22}$.}
\step{6}{Computes $U_{2j} = (X_{2j} - L_{21} U_{1j}) L_{22}^{-1}$ for $j = 3$ to $N$.}
\begin{itemize}[leftmargin=*, nosep]
  \item Forwards the received results from the previous server \\to $S_3$.
  \item Sends the computed values $U_{2j}$ (for $j = 2$ to $N$) to $S_3$.
  \item Returns result $\mathrm{res}_2 = \{L_{21}, L_{22}, U_{2j}\}$ (for $j = 2$ to $N$) \\to $C$.
\end{itemize}
\medskip

\textbf{$S_i$ Computation:}
\step{7}{Computes $L_{i1}, L_{i2}, \dots, L_{i(i-1)}$:
\[
L_{ik} = U_{kk}^{-1} \left( X_{ik} - \sum_{m=1}^{k-1} L_{im} \, U_{mk} \right),\]\\
\[\forall k = 1, \dots, i-1.\]}
\step{8}{Updates $X_{ii}$:
\[
X_{ii} = X_{ii} - \sum_{k=1}^{i-1} L_{ik} \, U_{ik}.
\]}
\step{9}{Performs $X_{ii} = L_{ii} \times U_{ii}$.}
\step{10}{Computes $U_{ij}$ for $j = i+1$ to $N$:
\[
U_{ij} = L_{ii}^{-1} \left( X_{ij} - \sum_{k=1}^{i-1} L_{ik} \, U_{kj} \right).
\]}
\begin{itemize}[leftmargin=*, nosep]
  \item Forwards the received results from the previous \\server along with the computed $U_{ij}$ to $S_{i+1}$.
  \item Returns result $\mathrm{res}_i$,\\ 
  $\mathrm{res}_i= \{L_{i1}, \dots, L_{ii}, U_{ii}, U_{i(i+1)}, \dots, U_{iN}\}$ to $C$.
\end{itemize}
\medskip
\medskip

\textbf{$S_N$ Computation:}
\step{11}{Computes $L_{N1}, L_{N2}, \dots, L_{N(N-1)}$:
\[
L_{Nk} = U_{kk}^{-1} \left( X_{Nk} - \sum_{m=1}^{k-1} L_{Nm} \, U_{mk} \right),\]\\
\[\forall k = 1, \dots, N-1.\]}
\step{12}{Updates $X_{NN}$:
\[
X_{NN} = X_{NN} - \sum_{k=1}^{N-1} L_{Nk} \, U_{kN}.
\]}

\step{13}{Performs $X_{NN} = L_{NN} \times U_{NN}$.}
\begin{itemize}[leftmargin=*, nosep]
  \item Returns result $\mathrm{res}_N = \{L_{N1}, L_{N2}, \dots, L_{NN}, U_{NN}\}$ \\to $C$.
\end{itemize}
\medskip

\textbf{Final Output Integration:}
\step{14}{Client $C$ integrates $\mathrm{res}_1, \dots, \mathrm{res}_N$ into $L$ and $U$.}

\vspace{6pt}\hrule height 0.8pt
\medskip
\medskip
The integrated representation of the results $\mathrm{res}_1, \dots, \mathrm{res}_N$ is
given by:
\[
\mathbf{L} =
\begin{bmatrix}
L_{11} & 0      & \cdots & 0 \\
L_{21} & L_{22} & \cdots & 0 \\
\vdots & \vdots & \ddots & \vdots \\
L_{N1} & L_{N2} & \cdots & L_{NN}
\end{bmatrix}
\]
and
\[
\mathbf{U} =
\begin{bmatrix}
U_{11} & \cdots & U_{1(N-1)} & U_{1N} \\
0      & \cdots & U_{2(N-1)} & U_{2N} \\
\vdots & \ddots & \vdots     & \vdots \\
0      & \cdots & 0          & U_{NN}
\end{bmatrix}.
\]

\medskip
\subsubsection{Inter-Server Communication Pattern}
The proposed parallel LU decomposition algorithm is designed to minimize idle time and reduce communication overhead between servers. In all configurations, in earlier multi-server models, communication follows a one-way result transfer pattern, in which each server transmits its computed results (i.e., $U_{ij}$) to the immediate downstream server. In earlier multi-server models, each server was required to broadcast its results to all downstream servers, causing significant delays as downstream servers had to wait for data from multiple upstream servers before initiating their computations.

In contrast, our proposed multi-server model improves efficiency by ensuring that each server activates its immediate successor after completing its first task in the case of $S_1$ to $S_2$, and after completing its first two tasks in all other cases (e.g., $S_i$ to $S_{i+1}$, where $i > 1$). Once these initial tasks are completed, the server sends the necessary results to the next downstream server, allowing it to smoothly begin its computations while the current server continues to compute and send further results in parallel. This staggered communication pattern reduces waiting time for downstream servers, ensuring that computations proceed in a parallelized manner, thereby maintaining full utilization of resources.

\subsection{RESULT AUTHENTICATION ALGORITHM (AUTHENTICATE)}
\noindent \textbf{Authenticate}$(L,U,\mathbf{X}) \rightarrow\{1,0\}$. The Authenticate algorithm ensures the correctness of the LU decomposition while maintaining computational efficiency. During the LU decomposition of the submatrices of $\mathbf{X}$, the corresponding blocks $L_{ij}$ and $U_{ij}$ are combined to form the full $\mathbf{L}$ and $\mathbf{U}$ matrices of size $n \times n$. When reconstructing $X'$ as $\mathbf{L} \times \mathbf{U}$, the reconstructed matrix $X'$ may match $\mathbf{X}$ exactly on an element-wise basis, but the determinant of $\mathbf{X}$ remains equal to or approximately equal. This discrepancy arises from both floating-point inconsistencies and the design of the parallel computing algorithm. While determinant comparisons could theoretically serve as a correctness check, computing the determinant of large matrices is computationally expensive, particularly on resource-constrained edge devices, making this method impractical in these scenarios.

To overcome this, we use lightweight authentication to avoid matrix-matrix multiplications. Gao et al.~\cite{Gao2023} proposed an authentication formula, which we denote as $Q_1$, that reduces matrix-matrix operations to matrix-vector operations using a random vector $\mathbf{r}$. We introduce $Q_2$ and $Q_3$ for greater efficiency in parallel computing: $Q_2$ uses a random vector, while $Q_3$ works directly on diagonal elements, eliminating the need for randomization. Unlike $Q_1$, both $Q_2$ and $Q_3$ return scalar values rather than vector output, making authentication more efficient.

\begin{figure*}
        \centering
        \includegraphics[width=0.9\textwidth, height=0.9\textwidth]{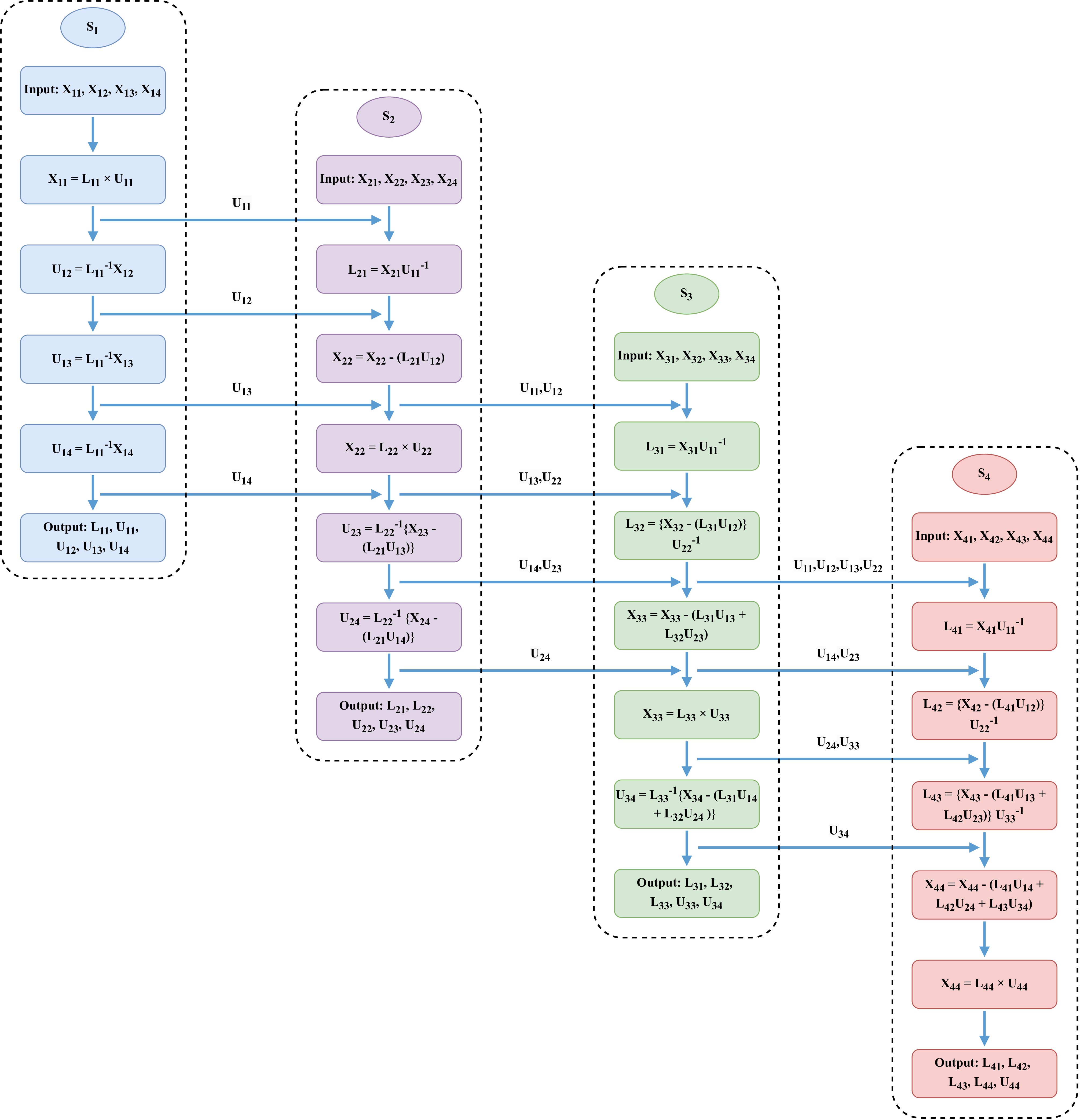} 
        \caption{Four-server parallel LU decomposition process.}
        \label{fig:6}
\end{figure*}

\subsubsection{Existing Formula: \(Q_1\)}
\centerline{$Q_1 = L \times (U \times \mathbf{r}) - \mathbf{X} \mathbf{r}$}
Produces a vector output of size $1 \times n$. In an ideal scenario, if $Q_1 = (0,0,\dots,0)^T$, the authentication succeeds (i.e., decomposition is correct). Otherwise, the result is rejected.

\subsubsection{Our Contributions: \(Q_2\) and \(Q_3\)}
We introduce two new authentication formulas that return scalar outputs, making them computationally lighter and easier to handle than the vector output of $Q_1$.

\medskip
\textbf{$Q_2$ – Scalar-Based Random Vector Authentication:}

\[
Q_2 = (L^{\mathsf{T}} \mathbf{r})^{\mathsf{T}} \times (U \mathbf{r}) - (\mathbf{r}^{\mathsf{T}} \times \mathbf{X}) \mathbf{r}
\]

Relies on a random vector $\mathbf{r} \in \mathbb{R}^n$, but it returns a scalar result.
\medskip

\textbf{$Q_3$ – Deterministic Scalar-Based Authentication:}
\[
Q_3 = \left| \sum_{i=1}^{n} \sum_{j=1}^{i} (L_{ij} U_{ji}) - x_{ii} \right|
\]

Relies solely on the deterministic values of the diagonal elements in $L$, $U$, and $\mathbf{X}$. It offers a quick and effective verification without requiring randomization.

\medskip
For both $Q_2$ and $Q_3$, in 2-server implementations, they return zero when the decomposition is correct. In multi-server cases, achieving zero is challenging, so a threshold $\epsilon$ is introduced to account for these deviations.

\subsubsection{Introduction of Threshold \(\epsilon\)}\label{sec:threshold}
A threshold $\epsilon(N)$ is introduced to account for numerical deviations that arise due to design limitations of the parallel computing algorithm and floating-point precision limitations. As $N$ increases, $\epsilon(N)$ adapts to compensate for discrepancies between $\mathbf{X}$ and $X'$, ensuring reliable authentication. The decomposition is validated if $\lvert Q \rvert < \epsilon(N)$, returning \texttt{Authenticate} = 1; otherwise, it is rejected with \texttt{Authenticate} = 0.

\subsection{RESULT DECRYPTION ALGORITHM (DECIPHER)}
\noindent \textbf{Decipher}$(\Psi,L,U) \rightarrow \det(\mathbf{M})$.
Upon successful authentication, the client decrypts the determinant of the original matrix $\mathbf{M}$ without relying on the secret key $K$, reducing security risks of $K$ being exposed. Instead, a seed-based correction factor $\Psi$ is used.

\subsubsection{Determinant Computation of Ciphertext Matrix \(X\):}
The determinant of the matrix $\mathbf{X}$ is computed from the diagonal elements of the matrices $L$ and $U$:
\[
\det(\mathbf{X}) = \prod_{i=1}^{n} \left( L_{ii} \times U_{ii} \right)
\]

\subsubsection{Determinant Recovery for Original Matrix \(M\):}
Depending on the obfuscation method used during encryption, $\mathbf{M}$'s determinant is retrieved as follows:
\subsubsubsection{For EWD + PRT:}
\[
\text{res} = \det(\mathbf{X}) \times (-1)^{\text{Rotate}(\Psi)} \times \text{Seed}(\Psi)
\]

\subsubsubsection{For EWM + PRT:}
\[
\text{res} = \frac{\det(\mathbf{X}) \times (-1)^{\text{Rotate}(\Psi)}}{\text{Seed}(\Psi)}
\]

The final result \textit{res} denotes the determinant of the plaintext matrix $\mathbf{M}$. By employing the seed-based correction factor during the deciphering process, this approach ensures determinant integrity while mitigating risks associated with key exposure.

\section{THEORETICAL ANALYSIS}
This section establishes the theoretical foundations of the proposed SPDC framework, covering Integrity, Privacy, and Authenticity through mathematical proofs and practical considerations. This demonstrates the reliability, scalability, and security of the framework.

\subsection{INTEGRITY}
The protocol ensures the client obtains the correct determinant of the original matrix $\mathbf{M}$, despite computations being performed on its encrypted counterpart $\mathbf{X}$. This integrity is maintained through secure partitioning, parallel LU decomposition, and result authentication.

\subsubsection{Parallel LU Decomposition}
After encryption, the matrix $\mathbf{M}$ is transformed into the matrix $\mathbf{X}$, which is divided into submatrices and distributed across $N$ edge servers for parallel LU decomposition. Each server $S_i$ begins the computation on its assigned submatrices and incrementally incorporates intermediate results received from upstream servers $S_{1}, \ldots, S_{i-1}$, thereby ensuring both consistency and integrity throughout the computation.

\[
X_{ii} \leftarrow X_{ii} - \sum_{j=1}^{i-1} L_{ij} U_{ji}
\]

Only after receiving these updates, server $S_i$ can proceed with its own LU decomposition:
\[
X_{ii} = L_{ii} U_{ii}
\]

This ensures that all computations integrate previous results, maintaining the correctness of LU factorization.

\subsubsection{Determinant Computation}
After receiving $L$ and $U$, the client reconstructs $\mathbf{X}$ and computes the determinant:
\[
\det(\mathbf{X}) = \prod_{i=1}^{n} L_{ii} \times \prod_{i=1}^{n} U_{ii} = \prod_{i=1}^{n} (L_{ii} \times U_{ii})
\]
The determinant of $\mathbf{M}$ is then recovered based on the applied EWO transformation:

\begin{itemize}
    \item{For EWD:}
    \[
        \det(\mathbf{M}) = \det(\mathbf{X}) \times (-1)^{\text{Rotate}(\Psi)} \times \text{Seed}(\Psi)
    \]
    \item{For EWM:}
    \[
        \det(\mathbf{M}) = \frac{\det(\mathbf{X}) \times (-1)^{\text{Rotate}(\Psi)}}{\text{Seed}(\Psi)}
    \]
\end{itemize}

Thus, if all servers adhere to the protocol, the determinant remains correct and tamper-resistant.

\subsection{PRIVACY}
The framework ensures input and output privacy through CED to protect both matrix elements and determinants.

\subsubsection{Input Privacy}
Input privacy is preserved through the combined use of two key techniques: EWO and the PRT-based obfuscation. Together, these transformations obfuscate both the values and structural arrangement of $\mathbf{M}$, resulting in $\mathbf{X}$, which is computationally indistinguishable from a random matrix.

\subsubsubsection{Element-wise Obfuscation (EWO) via Blinding Vector:}

Each element $m_{ij}$ of $\mathbf{M}$ is obfuscated as:

\begin{itemize}
    \item{EWD: } $m'_{ij} = \dfrac{m_{ij}}{v_i}$
    \item{EWM: } $m'_{ij} = m_{ij} \times v_i$
\end{itemize}

Here, $\mathbf{v} = [v_1, v_2, \dots, v_n]$ is a secret blinding vector derived from $\Psi$, ensuring that $\mathbf{X}$ remains computationally indistinguishable from random noise. Even if an adversary obtains full access to $\mathbf{X}$, they cannot reconstruct $\mathbf{M}$ without knowing $v_i$ for each element, as reversing EWD or EWM operations requires exact knowledge of $v_i$ in each transformation. Since $\mathbf{v}$ is derived from the cryptographic seed $\Psi$, which is kept private, the probability of successfully guessing $\mathbf{v}$ (and thereby reversing the transformations) is negligible.

\subsubsubsection{Structural Obfuscation via Panth Rotation Theorem (PRT):}
To further protect matrix structure, elements of matrix $\mathbf{M}$ are simultaneously rotated by a fixed rotation angle based on the quantized seed $\Psi'$. The rotation angle is computed based on the function $\text{Rotate}(\Psi)$, defined as:

\[
\text{Rotate}(\Psi) =
\begin{cases}
90^\circ & \text{if } (\Psi' \bmod 3) + 1 = 1 \\
180^\circ & \text{if } (\Psi' \bmod 3) + 1 = 2 \\
270^\circ & \text{if } (\Psi' \bmod 3) + 1 = 3
\end{cases}
\]

The rotated matrix, denoted as $M_R$, scrambles the spatial arrangement of the matrix elements without altering the matrix's determinant. Mathematically, for matrix $\mathbf{M}$, we represent this rotation as:

\[
\begin{aligned}
\mathbf{M}_R &= R_\theta(\mathbf{M}) \\
\det(\mathbf{M}_R) &= \det(\mathbf{M}) \times (-1)^{\text{Rotate}(\Psi)}
\end{aligned}
\]

where $\theta \in \{90^\circ, 180^\circ, 270^\circ\}$ is determined by the seed $\Psi$. This prevents pattern leakage and guarantees confidentiality, since even with access to $\mathbf{X}$, $L$, or $U$, an adversary cannot reconstruct $\mathbf{M}$ without knowledge of $\Psi$.

\subsubsection{Output Privacy}
Output privacy guarantees that the adversary cannot compute the true determinant $\det(\mathbf{M})$ from the outsourced computation results, even if they gain access to intermediate values (i.e., $L_{ij}$ and $U_{ij}$) or the fully encrypted matrix $\mathbf{X}$. It is preserved by the dependency of the determinant recovery formula on the seed $\Psi$ and the obfuscation method applied:

\begin{itemize}
    \item If EWD was applied:
    \[
    \det(\mathbf{M}) = \det(\mathbf{X}) \times (-1)^{\text{Rotate}(\Psi)} \times \text{Seed}(\Psi)
    \]
    \item If EWM was applied:
    \[
    \det(\mathbf{M}) = \frac{\det(\mathbf{X}) \times (-1)^{\text{Rotate}(\Psi)}}{\text{Seed}(\Psi)}
    \]
\end{itemize}

Seed$(\Psi)$ is derived from the security parameter and mathematical properties of $\mathbf{M}$. It is generated from a cryptographic source, ensuring that any attempt to infer $\det(\mathbf{M})$ remains infeasible under standard cryptographic assumptions.

\medskip
To further ensure privacy, the one-way inter-server communication pattern, where an upstream server passes intermediate $U_{ij}$ results only to its adjacent downstream server, restricts each server to partial computations. This prevents any single entity from reconstructing $\mathbf{X}$ or computing $\det(\mathbf{M})$. Even in the event of collusion and access to $\mathbf{X}$, servers lack access to the blinding vector and $\Psi$, making it impossible to recover $\mathbf{M}$ or $\det(\mathbf{M})$. This resilience to collusion is mathematically characterized by the information-theoretic guarantees provided by the protocol.

\medskip
For a probabilistic polynomial-time (PPT) adversary $A$ with access to $\mathbf{X}$, $L$, and $U$, the probability of successfully computing $\det(\mathbf{M})$ remains negligible:
\[
\Pr[A(\det(\mathbf{X}), L, U) = \det(\mathbf{M})] \leq \text{negl}(\lambda)
\]
where $\lambda$ is the security parameter tied to $\text{seed}(\Psi)$. This guarantees that cryptographic obfuscation and structured scrambling effectively safeguard both input and output privacy under adversarial conditions.

\subsection{AUTHENTICITY}
To ensure the correctness of LU decomposition in a multi-server environment, we analyze the mathematical foundations of the proposed $Q_2$ and $Q_3$ authentication methods, proving their reliability in verifying computational integrity.

\subsubsection{Proof of Correctness for \(Q_2\)}
The $Q_2$ authentication formula uses a random vector $\mathbf{r} \in \mathbb{R}^n$ to validate LU decomposition:
\[
Q_2 = (L^{\top}\mathbf{r})^{\top} (U \mathbf{r}) - (\mathbf{r}^{\top} X) \mathbf{r}
\]
For an accurate LU decomposition, $X = L \times U$. Substituting this into $Q_2$:
\[
Q_2 = ((L \mathbf{r})^{\top}(U \mathbf{r})) - (\mathbf{r}^{\top}(LU)\mathbf{r})
\]
Expanding both terms:
\[
((L \mathbf{r})^{\top}(U \mathbf{r})) = \mathbf{r}^{\top} LU \mathbf{r}, \quad \text{and} \quad (\mathbf{r}^{\top}(LU)\mathbf{r}) = \mathbf{r}^{\top} LU \mathbf{r}
\]
Since both expressions are equal, we obtain:
\[
Q_2 = 0
\]
Thus, $Q_2 = 0$ confirms the correctness of LU decomposition. If $Q_2 \neq 0$, it indicates computational errors or potential tampering, thereby signaling a failure in the decomposition.

\subsubsection{Proof of Correctness for $Q_3$}
The $Q_3$ authentication formula relies on diagonal consistency within $L$, $U$, and $\mathbf{X}$:
\[
Q_3 = \left| \sum_{i=1}^{n} \sum_{j=1}^{i} (L_{ij} U_{ji}) - x_{ii} \right|
\]

For a correctly decomposed matrix, each diagonal element $x_{ii}$ of $\mathbf{X}$ is given by:
\[
x_{ii} = L_{ii} \times U_{ii}
\]

\begin{table*}[t]
\centering
\caption{\textbf{Comparison of the Computational Overhead}}
\label{tab:1}
\begin{tabular}{lccccc}
\toprule
\textbf{Protocols} & \textit{SeedGen} & \textit{KeyGen} & \textit{Cipher} & \textit{Authenticate} & \textit{Decipher} \\
\mbox{[Reference]} & (biops) & (biops) & (flops) & (biops + flops) & (flops) \\
\midrule
Lei et al. \cite{Lei2015} & N/A & $(nm' + 2n + 3m')s$ & $2(n + m')^{2}$ & $l(n + m')s + 2l(n + m')^{2}$ & $4n + 5m'$ \\
Fu et al. \cite{Fu2017} & N/A & $(2nm' + n + 2m'^{2})s$ & $m'(n + m')^{2} + n^{2}$ & $l(n + m')s + 2l(n + m')^{2}$ & $3n + 2m'^{3} + 2m'$ \\
Liu et al. \cite{Liu2020} & N/A & $2ns$ & $4n^{2}$ & $lns + 2ln^{2}$ & $3n$ \\
Gao et al. \cite{Gao2023} & N/A & $ns$ & $2n^{2}$ & $ns + 2n^{2}$ & $3n$ \\
Our protocol & $2n$ & $ns$ & $n^{2}$ & $0 + 2n(n+1)$ ($Q_3$) & $2n$ \\
\bottomrule
\end{tabular}
\end{table*}

Substituting into $Q_3$:
\[
Q_3 = \sum_{i=1}^{n} \left| L_{ii} U_{ii} - x_{ii} \right| = 0
\]
Thus, $Q_3 = 0$ guarantees correct LU decomposition, while deviations indicate computational inconsistencies.

\subsubsection{Numerical Discrepancies in Multi-Server Implementations}  
According to Section~\hyperref[sec:threshold]{IV-E-3}, when the protocol is implemented across more than $2$ servers, slight numerical discrepancies arise primarily due to design limitations of the parallel computing algorithm and floating-point precision limitations. To accommodate this, authentication incorporates a threshold $\epsilon(N)$:
\[
|Q| \leq \epsilon(N) \quad \text{where } Q \in \{ Q_2, Q_3 \}
\]

Here, $\epsilon(N)$ increases as the number of servers $N$ grows, ensuring robust authentication without unnecessary rejection.

\section{PERFORMANCE EVALUATION}
This section presents a detailed performance evaluation of the proposed privacy-preserving and secure determinant computation protocol. Focusing on theoretical analysis rather than experimental testing, we assess computational complexity, communication cost, verification overhead, and deployment suitability. The results are benchmarked against state-of-the-art protocols \cite{Lei2015, Fu2017, Liu2020, Gao2023} to demonstrate comparative efficiency, scalability, and practicality.

\subsection{COMPUTATIONAL OVERHEAD}
Table~\ref{tab:1} compares the computational costs of the algorithms used in each protocol. Our proposed protocol demonstrates the lowest overall complexity across all stages.

Its streamlined encryption, authentication, and decryption processes significantly reduce the computational burden on both client and server. Unlike prior methods, it avoids redundant operations and multiple verification rounds under normal conditions, with additional computations required only when the server returns incorrect results. The parallel structure further enhances performance in large-scale, edge-oriented environments.

For fairness and consistency, comparative metrics for existing protocols are referenced from Gao and Yu \cite{Gao2023}. The overhead values, measured in floating-point operations (flops) and bitwise operations (biops), directly reflect actual resource consumption in high-performance computing contexts. Overall, our proposed framework emerges as the most scalable and practical solution for secure and privacy-preserving determinant computation, enabling secure outsourcing without imposing excessive load on clients or edge devices.

\subsection{PROTOCOL CHARACTERISTICS}
Table~\ref{tab:2} presents a comparative view of fundamental protocol characteristics, including privacy preservation, parallel outsourcing capability, and supported threat models. Early approaches such as \cite{Lei2015} lack both privacy protection and parallelism, while others like \cite{Fu2017, Liu2020} offer privacy guarantees but rely on serial computation or assume weaker adversaries. The recent work in \cite{Gao2023} advances toward parallel outsourcing under a malicious model, yet remains constrained to two servers and even-sized matrices.
\medskip

\begin{table}[h!]
\centering
\caption{\textbf{Protocol Characteristics Comparison}}
\label{tab:2}
\begin{tabular}{lccc}
\toprule
\textbf{Protocols} & \textit{Privacy} & \textit{Parallel} & \textit{Threat} \\
\mbox{[Reference]} & \textit{Preserving} & \textit{Outsourcing} & \textit{Model} \\
\midrule
Lei et al. \cite{Lei2015} & No & No & Malicious \\
Fu et al. \cite{Fu2017} & Yes & No & Honest-but-curious \\
Liu et al. \cite{Liu2020} & Yes & No & Malicious \\
Gao et al. \cite{Gao2023} & Yes & Yes & Malicious \\
Our protocol & Yes & Yes & Malicious \\
\bottomrule
\end{tabular}
\end{table}

In contrast, the proposed protocol uniquely integrates all three properties: privacy preservation, parallel outsourcing, and malicious adversary resistance, within a lightweight and scalable architecture. This is achieved through a combination of formal parameterization, modular algorithm design, and parallel execution strategies. The use of a sequential one-way communication pattern between servers helps reduce coordination overhead and simplifies parallel computation. Moreover, the protocol supports flexible deployment across any number of servers, including resource-constrained edge nodes. As a result, it enables efficient and secure determinant outsourcing for all square matrices, without the structural or trust limitations observed in prior work.

\subsection{MATRIX TYPE SUPPORT AND DIMENSION EXTENSION}
In MDC, handling both even- and odd-sized square matrices is critical. While determinants are defined for all square matrices, privacy-preserving protocols impose constraints that either disallow odd dimensions or require large padding to convert them into suitable even-sized matrices. Such differences directly impact computational efficiency and numerical accuracy.

\begin{table}[h!]
\centering
\caption{\textbf{Comparison of Protocols by Matrix Support and Extension}}
\label{tab:3}
\begin{tabular}{lcc}
\toprule
\textbf{Protocols} & \textit{Supported} & \textit{Requires Dimension} \\
\mbox{[Reference]} & \textit{Matrix Types} & \textit{Extension?} \\
\midrule
Lei et al. \cite{Lei2015} & Even \& Odd & Yes (always force-padded) \\
Fu et al. \cite{Fu2017} & Even \& Odd & Yes (always force-padded) \\
Liu et al. \cite{Liu2020} & Even \& Odd & No (padding not required) \\
Gao et al. \cite{Gao2023} & Even only & No (odd sizes not supported) \\
Our protocol & Even \& Odd & Yes (only when needed) \\
\bottomrule
\end{tabular}
\end{table}

Table~\ref{tab:3} compares representative protocols in terms of their ability to process even- and odd-sized square matrices and whether dimension extension is required. For instance, Lei et al. \cite{Lei2015} and Fu et al. \cite{Fu2017} enforce padding for all input sizes, while Gao et al. \cite{Gao2023} entirely exclude odd-sized matrices from consideration. Liu et al. \cite{Liu2020} avoid padding, directly supporting both cases. Our protocol, in contrast, supports both even and odd dimensions, applying minimal padding only when strictly necessary.

\begin{table}[h!]
\centering
\caption{\textbf{Comparison of Protocols by Dimension Extension Mechanism}}
\label{tab:4}
\begin{tabular}{lcc}
\toprule
\textbf{Protocols} & \textit{Dimension Extension Mechanism} \\
\midrule
Lei et al. \cite{Lei2015} & Always extends to $(n+m') \times (n+m')$ \\
& via block augmentation \\[0.3ex]
Fu et al. \cite{Fu2017} & Always extends to $(n+m') \times (n+m')$ \\
& via block augmentation \\[0.3ex]
Liu et al. \cite{Liu2020} & Works directly on original $n \times n$ matrix \\[0.3ex]
Gao et al. \cite{Gao2023} & Only partitions even-sized matrices \\
& into blocks for parallel processing \\[0.3ex]
Our protocol & Pads odd $n$ to next even; can optionally\\
& pad even $n$ as per server requirements \\
\bottomrule
\end{tabular}
\end{table}

Table~\ref{tab:4} details the mechanisms by which protocols extend matrix dimensions when required. Most prior methods rely on block augmentation, which inflates matrix size and may introduce redundant computation. Our protocol applies padding only when necessary and follows a determinant-preserving augmentation scheme. It applies padding in two distinct scenarios. First, it uses nearest-even padding to minimally extend odd-sized matrices to the next even order. Second, when the user opts for parallel computation across many servers to accelerate determinant evaluation, the protocol pads even-sized matrices if the input is too small to be evenly partitioned across all N servers, ensuring balanced encrypted block distribution. This design achieves a balance between computational efficiency and correctness, ensuring result fidelity across diverse matrix types and deployment scenarios.

\subsection{DEPLOYMENT AND CLIENT COMPATIBILITY}
As detailed in Table~\ref{tab:5}, earlier protocols were primarily designed for cloud-centric environments, where resource limitations on participating devices are typically not a primary concern. While the recent work in~\cite{Gao2023} introduces an edge-oriented approach, it remains restricted to a fixed $2$-server model and offers only limited support for deployment on heterogeneous or resource-constrained platforms. In contrast, our protocol is explicitly developed to operate efficiently across an arbitrary number of untrusted servers and is fully compatible with edge devices that have limited computational capacity, making it more suitable for practical real-world deployments. 

\begin{table*}[t]
\centering
\caption{\textbf{Deployment-Oriented Properties of Determinant Outsourcing Protocols}}
\label{tab:5}
\begin{tabular}{lccccc}
\toprule
\textbf{Protocols} & \textit{Deployment} & \textit{Server} & \textit{Verification} & \textit{Result Extraction} & \textit{Resource-Constrained} \\
\mbox{[Reference]} & \textit{Environment} & \textit{Scalability} & \textit{Method} & \textit{Mechanism} & \textit{Support} \\
\midrule
Lei et al. \cite{Lei2015} & Cloud & $1$ (single cloud server) & Multi-round Monte Carlo & Key-based & No \\

Fu et al. \cite{Fu2017}   & Cloud & $1$ (single cloud server) & Multi-round Monte Carlo & Key-based & No \\

Liu et al. \cite{Liu2020} & Cloud & $1$ (single cloud server) & Multi-round Monte Carlo & Key-based & Partial \\

Gao et al. \cite{Gao2023} & Edge & $2$ (fixed block partition) & $Q_1$: Single-round prob. (vector) & Key-based & Partial \\

Our protocol  & Edge & $N$ (flexible, arbitrary) & Single-round (scalar): & Seed-based & Yes \\

  &  &  & $Q_2$ prob., $Q_3$ deterministic &  &  \\
\bottomrule
\end{tabular}

\vspace{2mm}
\centering
\parbox{0.9\textwidth}{\small \textit{Our protocol is designed for edge environments, supporting flexible parallelism with arbitrary $N$. 
It can adapt to federated or cloud clusters, though such deployment is not the primary focus of our analysis.}}

\end{table*}

To improve efficiency and security, the protocol avoids conventional designs that depend on multi-round probabilistic verification and key-based result extraction. It instead uses a single round of verification, which may be either deterministic ($Q_3$) or probabilistic ($Q_2$), and adopts a seed-based result extraction mechanism. This design not only significantly reduces computational and communication overhead, but also lowers the associated security risks by minimizing sensitive key usage on the client side. Together, these features enable scalable, lightweight, and secure determinant outsourcing in diverse edge computing environments.

\section{EXTENSIONS AND APPLICATIONS}
The proposed framework is designed for scalability and adaptability, allowing it to be applied across diverse computational environments. This section explores its current applications and potential future advancements in various domains, emphasizing its versatility and effectiveness.

\subsection{SCALABILITY AND SECURE COMPUTATION IN DISTRIBUTED ENVIRONMENTS}
Our SPDC framework supports multi-server scalability and decentralized computation, making it adaptable to cloud computing, edge computing, and IoT networks.
\begin{itemize}
    \item \textbf{Parallel Computation Efficiency:} The SPCP protocol reduces computational complexity as $N$ increases, ensuring scalability for large-scale matrix determinant computations.
    \item \textbf{Client-Controlled Multi-Server Deployment:} Unlike fixed $2$-server models, the privacy-preserving computation (PPC) protocol allows clients to allocate and manage any number of servers, dynamically partitioning matrices for parallel execution.
    \item \textbf{Secure IoT and Ad Hoc Computing:} The protocol enables distributed determinant computation in resource-constrained environments, such as IoT and ad hoc networks, where computational tasks are shared across multiple devices while preserving data privacy.
\end{itemize}

\begin{figure}[ht!]
    \centering
    \begin{minipage}{0.48\textwidth} 
        \centering
        \includegraphics[width=\textwidth,height=7cm,keepaspectratio]{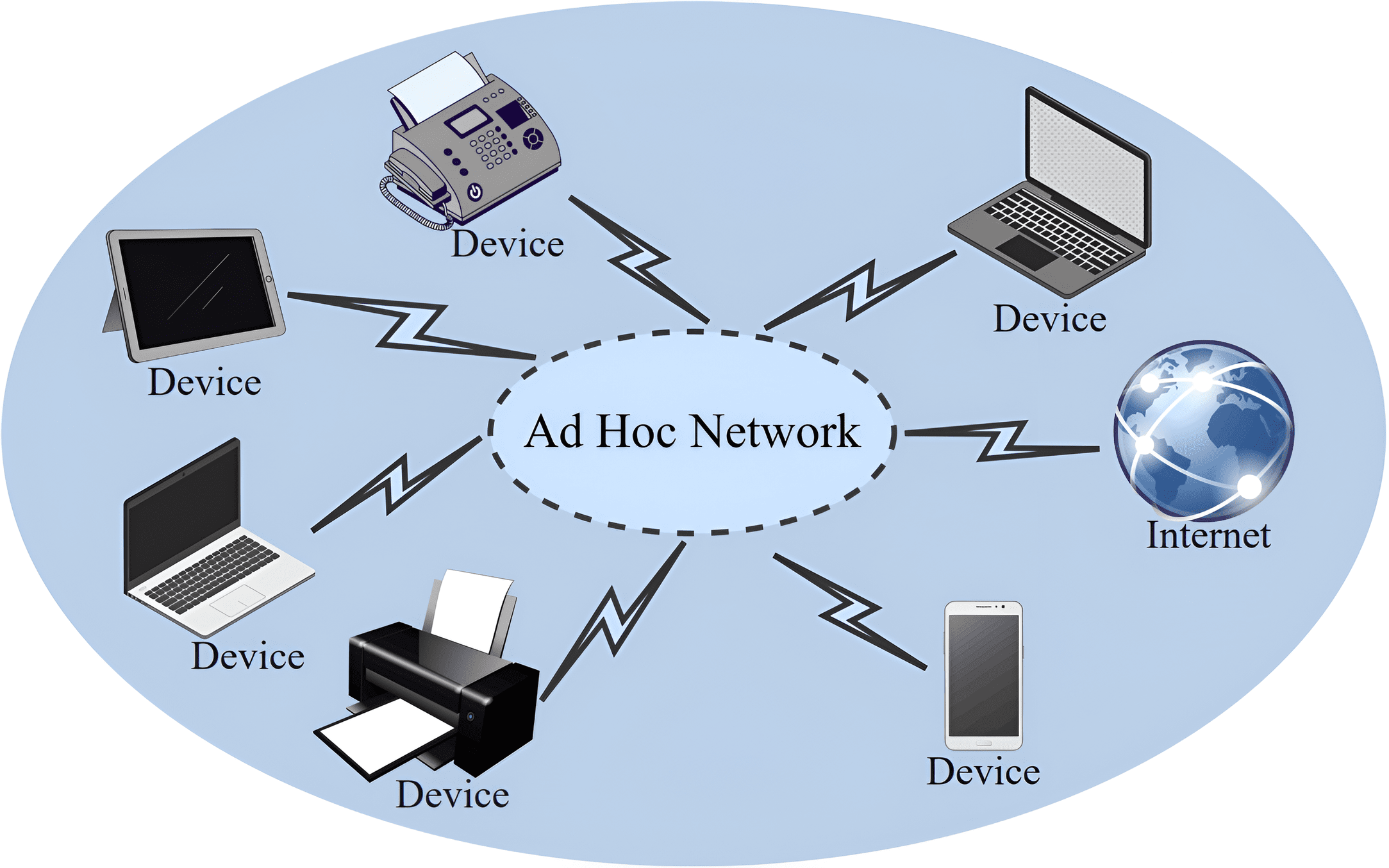}
        \vspace{-0.1mm}
    \end{minipage}
    \caption{Ad hoc network model.}
    \label{fig:7}
\end{figure}

\subsection{FUTURE ENHANCEMENTS AND EXTENSIONS}
While the SPDC framework effectively supports secure $N$-server determinant computation, it has the potential to support broader privacy-preserving and secure matrix operations and integrated authentication mechanisms.
\begin{itemize}
    \item \textbf{Blockchain-Integrated Secure Computing:} Integrating blockchain into the PPC protocol can enhance authenticity by providing immutable audit trails and consensus mechanisms. This ensures transparent, verifiable computations across untrusted servers, strengthening security against adversarial attacks \cite{Nakamoto2008}.
    
    \item \textbf{Privacy-Preserving Matrix Inversion \& Eigenvalue Computation:} Extending the protocol to support secure matrix inversion and eigenvalue computation would open applications in machine learning, optimization, and secure data analysis.
    
    \item \textbf{Automated Fault Tolerance:} Future implementations could incorporate real-time failure detection, redundancy strategies, and dynamic task redistribution, ensuring uninterrupted computation in cloud, edge, and MANET environments.
\end{itemize}

This framework provides a scalable, lightweight, and reliable $N$-server determinant computation protocol with applications in distributed computing, IoT security, and real-time analytics. Future advancements in blockchain authentication, expanded matrix operations, and automated fault tolerance will further enhance its impact across scientific, industrial, and AI-driven domains.

\section{CONCLUSION}
This paper presented the Secure Parallel Determinant Computation (SPDC) framework, a privacy-preserving and secure approach to parallel matrix determinant computation tailored for resource-constrained and distributed edge environments. The framework employs Composite Element Distortion (CED), which integrates Element-wise Obfuscation (EWO) with Panth Rotation Theorem (PRT)-based Obfuscation. Through this mechanism, SPDC enables scalable $N$-server parallel LU decomposition over untrusted edge servers while ensuring strong data confidentiality, computational integrity, and high efficiency.

Performance evaluations show that the proposed method significantly reduces computational overhead on both client and server sides, eliminates unnecessary dimension extension, and supports both even and odd-sized square matrices via minimal, determinant-preserving augmentation. The use of lightweight, single-round, scalar-based authentication mechanisms enables rapid and robust verification without burdening edge devices.

The framework’s adaptable design enables secure and parallel outsourcing across heterogeneous infrastructures, thereby overcoming the scalability, privacy, and deployment limitations of prior solutions. Collectively, these contributions establish a new benchmark for practical, efficient, and secure determinant outsourcing in real-world Internet of Things (IoT) and large-scale distributed computing scenarios.

\bibliographystyle{IEEEtran}
\bibliography{references}

\begin{IEEEbiography}[{\includegraphics[width=1in,height=1.25in,clip,keepaspectratio]{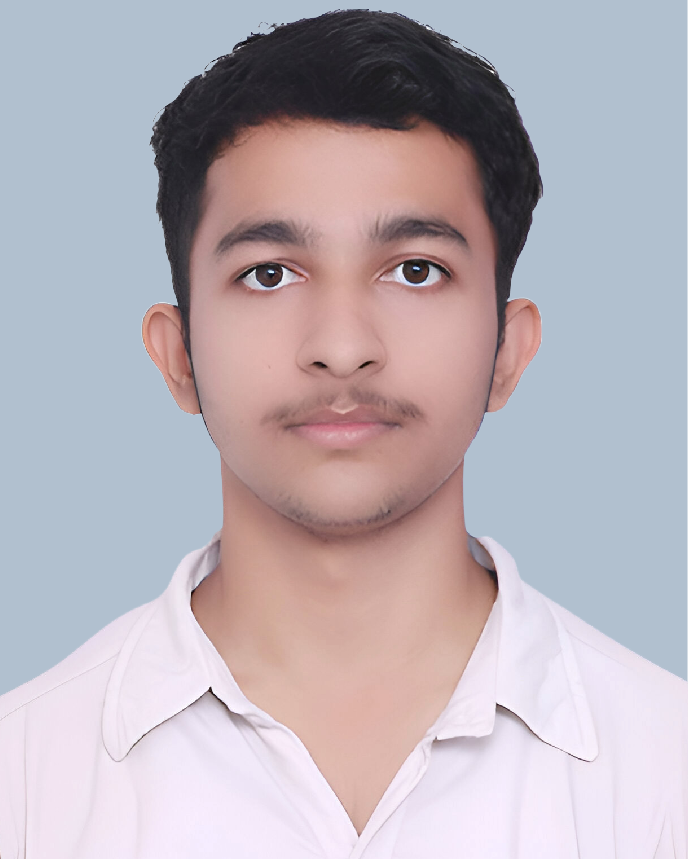}}]
{Prajwal Panth}~(Member, IEEE)~is currently pursuing a B.Tech. (Honours with Research) in Computer Science and Engineering, with a minor in Applied Machine Learning, at the Kalinga Institute of Industrial Technology (KIIT), India. His research interests include machine learning, secure computation, secure outsourced computation, and natural language processing. He received the Scientific Innovation Award from the Government of Nepal in recognition of his contributions to research and science outreach.
\end{IEEEbiography}

\vfill\pagebreak

\end{document}